\documentclass[draft]{agujournal2019}
\usepackage{url} 
\usepackage{lineno}
\usepackage[inline]{trackchanges} 
\usepackage{soul}
\usepackage{amsmath}
\usepackage{amssymb}
\usepackage{minted}
\journalname{JAMES}

\newcommand{\andeq}[0]{\quad\text{and}\quad}
\newcommand{\bs}[1]{\boldsymbol{#1}} 

\newcommand{\beq}{\begin{equation}}
\newcommand{\eeq}{\end{equation}}
\newcommand{\bU}{\bs{U}}
\newcommand{\coh}{\widehat{C}}

\definecolor{HW}{RGB}{137,0,225}

\newcommand{\ch}[1]{#1}

\setstcolor{blue}

\newcommand{\chm}[1]{#1}

\begin{document}

%
%


\title{Disentangling Internal Tides from Balanced Motions with Deep Learning and Surface Field Synergy}

%
%




\authors{Han Wang\affil{1}, Jeffrey Uncu\affil{2}, Kaushik Srinivasan\affil{3}, Nicolas Grisouard\affil{2}}

\affiliation{1}{Institute of Oceanography, University of Hamburg, 20146 Hamburg, Germany}
\affiliation{2}{Department of Physics, University of Toronto, Toronto ON M5S\,1A7, Canada}
\affiliation{3}{University of California, Los Angeles, CA 90095, USA}





\correspondingauthor{Han Wang}{hannnwangus@gmail.com}



\begin{keypoints}
\item The imprints of incoherent internal tides and balanced motions on sea surface height are disentangled by a deep learning approach 
\item 
Surface velocity is by far the most \chm{informative input under our benchmark} \ch{while sea surface height and surface temperature are informative too}
\item \ch{Mesoscale context improves internal tide extraction and more local networks perform worse}
\end{keypoints}

%
%

%
%


\begin{abstract}
A fundamental challenge in ocean dynamics is the disentanglement of balanced motions and internal waves. Extracting internal tidal (IT) imprints on surface data is a central part of this challenge. For IT extraction,  traditional harmonic analysis fails in the presence of strong incoherence and poor temporal sampling, as is common in global satellite observations. The advent of new wide-swath satellites, which provide two-dimensional spatial coverage, allows IT extraction to be reformulated as an image translation problem.
Building on recent work where we developed a deep learning approach to extract IT signatures from sea surface height (SSH) in an idealized turbulent simulation, we show here that a simpler and computationally cheaper algorithm can perform \ch{comparably in our experiments} 
if the learning rate is annealed during training.
Using this new, convenient algorithm, we experiment with different combinations of input surface fields -- SSH, surface temperature, and surface velocity. All fields contribute synergistically to disentanglement \chm{in our deterministic benchmark}, with surface velocity by far the most informative. These findings underscore the value of coordinated multi-platform observational campaigns and highlight the critical importance of surface velocity observations for separating balanced motions and internal waves.  
Additional insights into the behavior of the deep learning  algorithm emerge: both wave-signature \ch{information }and scattering-medium \ch{information }aid IT extraction, and to exploit large-scale\ch{, mesoscale-reaching} information in the scattering medium, the algorithm must be highly non-local. Residual errors of our algorithm concentrate at small spatial scales near mode-2 tidal wavelengths, likely arising from
\chm{a combination of incomplete information content in the chosen inputs, uncertainty in the simulation-derived reference fields (e.g., Doppler-shift-related contamination), and limitations of the present deterministic deep-learning architecture.}
\end{abstract}

\section*{Plain Language Summary}
The ocean hosts motions across a wide range of time and space scales. Some are long-lived, such as currents and eddies, while others are waves, such as internal tides generated when tidal currents flow over bottom topographic features. These different motions are very hard to separate. A new NASA satellite, SWOT, can measure the sea surface at fine spatial resolution, but because it passes the same place only every few weeks, traditional methods of extracting internal tides do not work well. We use machine learning to test whether combining different surface measurements---sea surface height, surface temperature, and surface currents---can help. Our results show that surface currents are the most important piece of information, but that combining all three fields works best. These findings support future satellite missions designed to measure currents directly, and/or with concurrent measurements of different surface fields, which will improve our ability to track how tides move energy through the ocean. We also gain insights on the behavior of the machine learning algorithm and understand conceptually how some design considerations work.  

\section{Introduction}
A fundamental challenge in ocean dynamics is the separation of balanced motions (hereafter ``BMs") and internal waves (hereafter ``waves").
BMs arise when time-derivative terms are small relative to other terms such as effects of rotation, stratification, and advection in the dynamical equations. Their time scales span weeks to months.
They include climate-scale circulations and submesoscale currents \cite{Mcwilliams2016}, taking up the bulk of the ocean's kinetic energy \cite{2009_ARFM_FerrariWunsch}. 
Also ubiquitous in the ocean are  waves, which are unbalanced motions.  
Their frequencies are typically above the Coriolis frequency, with time scales of hours. 
They include, among others, storm-forced near-inertial waves 
and internal tides (hereafter ``ITs") that are  generated as the astronomically forced tidal currents flow over underwater topography. 
BMs and waves have different impacts on tracer and energy transports, and an accurate BM-wave separation is crucial for representing dynamical processes like wave-driven mixing to create reliable maps of ocean circulation and to adequately represent the ocean's role in the climate \cite{garrett1972oceanic, polzin1997spatial,2009_ARFM_FerrariWunsch,whalen2020internal}.

This separation is difficult. Even with full temporal and spatial information (typically from simulations), BM-wave separation can be conceptually fraught and can only be made at the estimation level, as BMs can have overlapping temporal or spatial scales and strong dynamical interactions with waves \cite{vanneste2013balance, barkan2024eddy, kar2025spontaneous}. A common approach is to assume time scale separation: slow motions are interpreted as BMs, and fast motions as waves \cite{Savva2019, ShakespeareGibsonHogg2021,jones2023using}.
With observational data, where temporal/spatial information is limited, even this estimated separation becomes challenging. Global satellite observations of sea surface height (hereafter ``SSH”) are a classic example of this limitation: repeat cycles of tens of days are far longer than wave periods, so simple frequency filtering is ineffective. For such altimetry observations, progress has been made under a less ambitious objective: rather than separating the full wave spectrum from BMs, some efforts have focused on extracting ITs \cite{munk1966tidal,ray2011non,zhao2012mapping,carrere2021accuracy,WangGrisouard2022,zhao2024internal}, which constitute a significant portion of wave energy \cite{kantha1997global}. 
ITs are more trackable than other wave components: under given background conditions, they concentrate around predictable frequencies and wavenumbers---a fact that is the basis of the classic approach of harmonic fitting, which matches observations to plane-wave-based models. Meaningful global maps of ITs are thus produced \cite{zhao2012mapping, carrere2021accuracy}, illuminating mixing properties \cite{whalen2020internal} and climatological propagation properties  of ITs \cite{zhao2016internal}. 

This progress must be interpreted with a commonly acknowledged caveat: IT harmonic fitting is inaccurate for coarsely sampled data in presence of strong incoherence. Incoherence\ch{, in this context, refers to having phase shifts relative to the astronomical forcing that are not constant over the time window used for harmonic fitting of altimetry data; for satellite altimetry, such windows are commonly longer than 10 days. IT incoherence }is ubiquitous in ITs. As ITs propagate, their propagation angles and spatial scales are constantly modulated by the turbulent BM and varying background conditions. 
Linear superposition of scattered ITs alone produces time- and space-dependent changes in phase and wavenumber, even without additional wave–wave interactions.
In observations, 
IT incoherence is found universally across different types of data \cite{ray2011non, lob2020observations, caspar2025combining}. 
In simulations and analytical models, the causal effects of BMs and background conditions on IT incoherence are extensively demonstrated \cite{2014_JGR_DunphyLamb,2015_GRL_PonteKlein,kelly2016coupled,caspar2022characterization,lahaye2024internal,uncu2024wave,wang2025agulhas}.
The approach of harmonic fitting is developed with incoherence in mind  \cite{munk1966tidal, zhao2024internal} and can resolve incoherent ITs if sampling is sufficiently fine. But for coarsely sampled global satellite SSH, fine-scale incoherence cannot be reliably captured by standard plane-wave–based harmonic fitting schemes. \citeA{carrere2021accuracy} provides a quantified review of this issue. This can yield qualitatively incorrect results:
for example, \citeA{buijsman2017semidiurnal} argues that the apparent IT damping in equatorial Pacific, inferred via harmonic fitting of decades of satellite SSH data, may reflect amplified fitting errors due to increased incoherence rather than true IT energy loss. Such systematic error hinders an accurate global picture of IT-related energy pathways. 

The recently launched Surface Water and Ocean Topography (hereafter ``SWOT”) mission shares the long repeat cycles of earlier single-satellite altimeters (about 21 days, still much longer than tidal periods), so the limitations of harmonic fitting in the presence of incoherence persist. What SWOT fundamentally changes, however, is the spatial sampling: it delivers wide-swath SSH snapshots at unprecedented spatial resolution,  creating an opening for methods that can exploit spatial information.


New approaches have sought to incorporate additional dynamical constraints to compensate for the lack of temporal data. These physics-based approaches range from potential vorticity inversion that leverages concurrent surface density and SSH data to reconstruct the BM 
\cite{Ponte2017, wang2025practical},
to variational data assimilation that fit observations to coupled balanced and wave models \cite{bellemin2025variational}. 
Data-driven reduced-order methods have also emerged. For instance, 
Proper Orthogonal Decomposition based on prior knowledge of the BM's structure isolates the IT component that is correlated with the eddy field \cite{maingonnat2025coupled}. For cases where relatively short revisiting periods are available, for example during the initial SWOT Cal/Val track, Dynamical Mode Decomposition can effectively perform a spatio-temporal filter 
\cite{lapo2025method, uchida2025dynamic}. A comprehensive comparison of these methods, using robust benchmark data
from simulations (where spatio-temporal data is fully accessible), is still lacking.

In parallel, deep learning has emerged as a promising alternative. Recent works treat flow separation as an image-to-image translation problem as opposed to filtering in time, and train neural networks on large simulated datasets where the BM and waves are known. \ch{The earliest, proof-of-concept work in this direction we are aware of is \citeA{lguensat2020filtering}, which showed, for IT filtering, that a convolutional neural network can outperform a linear spatial filter.}
\citeA{WangGrisouard2022} (hereafter ``W22") used a conditional Generative Adversarial Network (hereafter ``cGAN") to extract ITs from SSH snapshots in an idealized simulation where a baroclinic jet interacts with a mode-1 IT. 
\ch{Other works address a broader wave--BM, snapshot-based decomposition problem than the IT extraction problem considered here, where the wave component in principle contains all wave motions resolved by simulations. 
An explicit attempt to address multi-scale fidelity is made in \citeA{lyu2024multi}, which introduced a ZCA-whitened training framework to reduce spectral bias in the wave--BM decomposition. \citeA{Wang2025MultiScale} extended the same framework to a probabilistic formulation. \citeA{liu2025wide} uses a U-Net-like model with Mamba blocks, plus a built-in frequency-separation module, also in an attempt to improve multi-scale fidelity. 
Some of these works target the BM rather than the waves:} \citeA{Gao2024} uses a U-Net with a loss function that heavily weights the SSH gradient error, ensuring the recovered BM has realistic geostrophic currents. Related but distinct, \ch{\citeA{xiao2023reconstruction} uses a U-Net trained with a standard pointwise loss to regress surface kinematic fields (vorticity, strain, and divergence) directly from SSH, showing that the network tends to suppress inertia-gravity-wave divergence without any explicit wave--BM decomposition or physics-informed loss term.} \ch{All works mentioned are  SWOT-motivated, with \citeA{Wang2025MultiScale,liu2025wide} already presenting proof-of-concept applications to SWOT observations. 
These studies collectively show that snapshot-based learning is promising, but they are not yet directly comparable quantitatively, as they differ in target outputs, benchmark data, input variables, and reported skill metrics.}

Here we expand the methodology of W22, which used solely SSH as input. We broaden the scope of possible input fields to include surface horizontal velocities (hereafter ``surface velocities") and sea surface temperature (hereafter ``SST").
\ch{In previous works,  
\citeA{lguensat2020filtering} and \citeA{Wang2025MultiScale} have studied the effect of adding SST to SSH snapshots, and \citeA{Wang2025MultiScale} reports a moderate improvement from doing so. This is broadly consistent with our result and interpretation that SST provides useful information on the wave-scattering medium. Our work differs as we visit the question in a different setting, and, more importantly, extend it by systematically comparing SSH, SST, and surface velocity as concurrent inputs.}
These additional input fields are in principle observable in the ocean. 
The SST is commonly derived from satellite observations using thermal infrared and passive microwave sensors, each with complementary characteristics, and  global gap-free products spanning decades of observations already exist; \citeA{nielsen2024impact} includes a recent review. 
Surface velocities have been traditionally measured by coastal high-frequency radars \cite{terrill2006data}, which already  overlap with SWOT's SSH observations \cite{kachelein2024characterizing}, and are demonstrably informative about incoherent ITs in the measured regions.
In the open ocean, extensive measurements of $\bs{U}$ are relatively novel, made possible by recent developments of remote sensing instruments and data retrieval algorithms \cite{hauser2023satellite}; mission concepts such as 
\ch{SEASTAR} \cite{mccann2024new} and HARMONY \cite{9554076} are currently under evaluation \ch{or} preparation; \ch{some} regional campaigns for evaluation and proofs of concept are completed \ch{\cite{mccann2024new}}.  
Our primary objective is to identify which fields provide the most valuable information \chm{in the present deterministic benchmark, }and how they act in synergy. The results clarify the contribution of each field, highlight the critical role of surface velocity, and offer valuable insights for optimizing future observational and data preparation strategies to best disentangle the ocean's complex, interacting balanced and unbalanced motions. 

To facilitate experiments, we streamline the deep learning algorithm, discovering that with some care in learning-rate scheduling, a U-Net achieves performance on par with the more complex cGAN used in W22. 
Using data from a well-established Boussinesq simulation \cite{2015_GRL_PonteKlein,Dunphy2017} that features a mode-1 IT propagating through a turbulent BM, we test all combinations of sea surface height ($H$), surface velocities ($\bU=(U,V)$), and surface temperature ($T$) as inputs. These are described in detail in \S \ref{sec:Unet:methodology}. 
We conduct a systematic analysis in \S \ref{sec:impacts} on how different, concurrently measured surface fields contribute to the deep-learning-based extraction of IT.
Along the way, we formulate a physics-motivated perspective explaining the different impacts from different input fields.  \ch{A spectral analysis of the performance shows that} residual errors concentrate at small spatial scales 
(\S \ref{sec:capturedmissed}). \ch{Additional experiments with degraded inputs show that the utility of using surface velocities or surface temperature as additional inputs is not limited to cases of perfectly simultaneous, high-resolution inputs (\S \ref{sec:degraded}).}
\ch{Furthermore, we find that the ability to leverage non-local, mesoscale-reaching information} is important for the algorithm's success (\S \ref{sec:nonlocal}). A summary and perspectives on  future developments are offered in \S \ref{sec:Unet:discussion}.

\section{Methodology} \label{sec:Unet:methodology}
\subsection{Boussinesq Simulation} \label{sec:BoussSim}
The training and testing data for our deep learning algorithms are outputs from an idealized simulation described in \citeA{2015_GRL_PonteKlein} and \citeA{Dunphy2017}. This set of simulations is developed by members of the SWOT Science Team with the challenge of incoherent ITs in mind, and well established as a complex but idealized benchmark for studies of IT-BM interactions and disentanglement \cite{Ponte2017,LeGuillou2021,caspar2022characterization}. 
The specific dataset we use is uploaded on \citeA{BoussData}. W22 used the same simulation data set. We briefly recapitulate some directly relevant aspects. 

\begin{figure}
    \centering
    \includegraphics[width=0.99\linewidth]{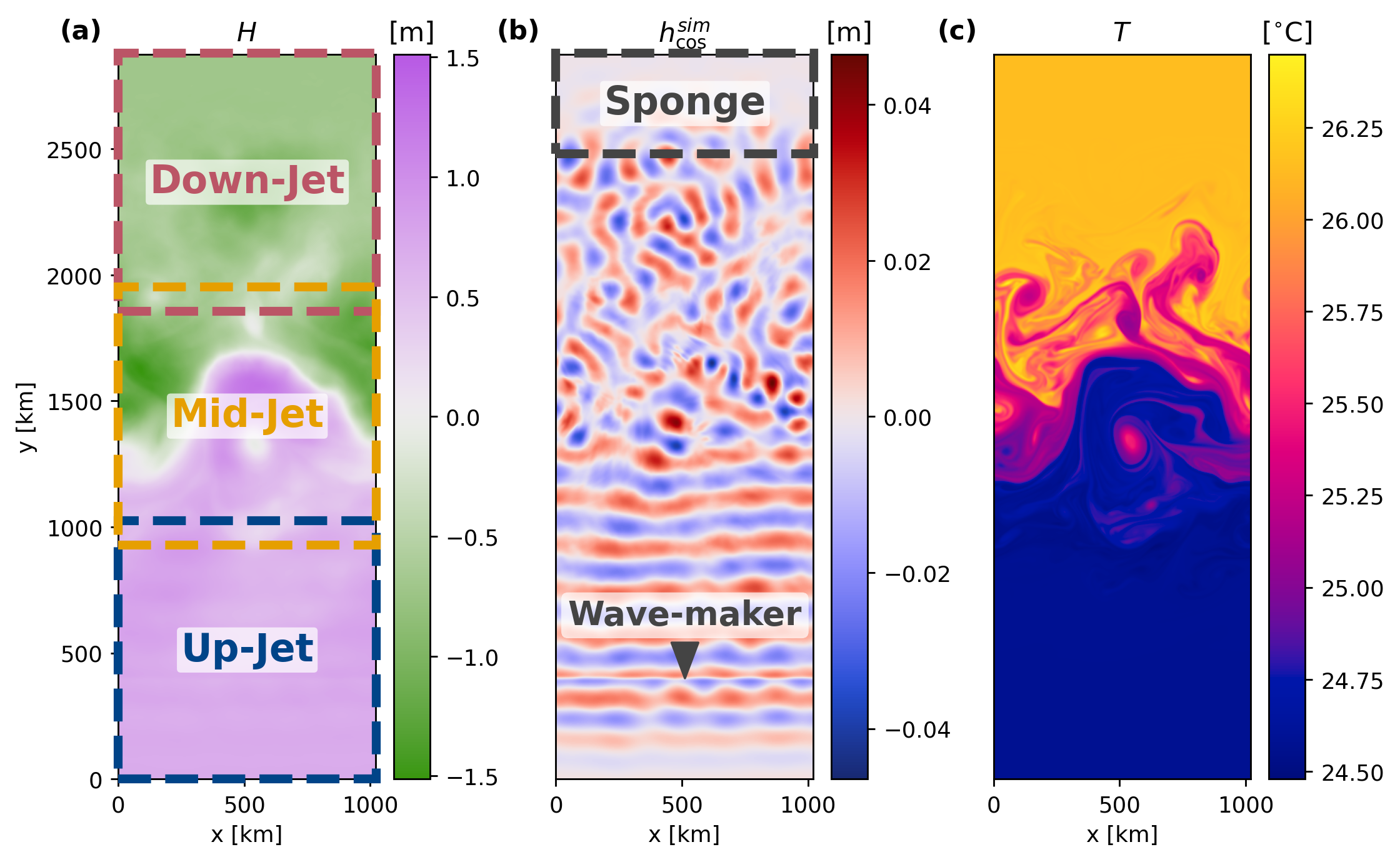}
    \caption{Snapshot of raw SSH $H$ (panel (a)), IT signal $h^{\text{sim}}_{\cos}$ (panel (b)), and surface temperature $T$ (panel (c)) from the T5 simulation at day 200. 
    Dashed boxes mark the up-jet, mid-jet, and down-jet regions defined for analysis. The Sponge region and the central latitude of the wave-maker are marked in panel (b).  }
    \label{fig:Bouss_setup}
\end{figure}

A three-dimensional, hydrostatic Boussinesq model is solved by the Regional Oceanic Modeling System \cite{ShchepetkinMcWilliams2005}, with horizontal resolution of $4$ km. Equations are integrated on a $\beta$-plane centered around latitude $45^\circ$ N. A snapshot of some surface fields plotted over the whole horizontal domain is shown in Fig.~\ref{fig:Bouss_setup}.
The model contains two essential dynamical components for our investigation: an incoherent IT, and a turbulent BM that scatters the IT. 
A continuously restored meridional density gradient in central latitudes induces baroclinic instability, forming a turbulent zonal jet 
with a characteristic meander width of $\sim 800$~km. 
This constitutes the turbulent BM. 
A wave-maker at the southern end of the domain radiates a mode-1 IT with a period of $
P=12$ hours. A northern sponge layer 
prevents reflections. The strong incoherence of ITs is induced by (a), the scatterings from the BM and (b), the modulations of peak wavelengths induced by spatially varying density profiles.
To distinguish the performance of our deep learning algorithms at different locations relative to the turbulent jet, we partition the domain into up-jet, mid-jet and down-jet regions marked in Fig.~\ref{fig:Bouss_setup}(a). The mid-jet region spans the latitudes with nonzero background meridional density gradients, and accordingly, the turbulent jet's kinetic energy is stronger in the mid-jet region than the up/down-jet regions. The ITs are plane-wave-like in the up-jet region, which is close to the forcing region and not yet reaching the regions with strong BMs. 
As the BMs scatter ITs and introduce incoherence, ITs become strongly incoherent mid-jet (Fig.~\ref{fig:Bouss_setup}, panel (b)). ITs in the down-jet region are strongly incoherent even though the BMs are not as active there; this is primarily because the BMs alter the directions of ITs, introducing non-local impacts downstream of the main propagation directions. 

The reference fields indicating IT imprints on SSH are defined following \citeA{2015_GRL_PonteKlein}:
\begin{eqnarray}
 h^{\text{sim}}_{\cos}(x, y, t) = \frac1P\int^{t}_{t-2P} H(x, y, t')\cos\left(\frac{2\pi t'}{P}\right) \text dt',&& \andeq 
 \label{eq:eularian filtering1}\\ 
 h^{\text{sim}}_{\sin}(x, y, t) = \frac1P\int^{t}_{t-2P} H(x,y, t')\sin\left(\frac{2\pi t'}{P}\right) \,\text dt',&&
 \label{eq:eularian filtering2}
\end{eqnarray}
where $t$ denotes time in the Eulerian frame, and the superscript ``sim" notes that the fields are produced by the Boussinesq ``simulation" data. 
The components $h^{\text{sim}}_{\cos}$ and $h^{\text{sim}}_{\sin}$  constitute the reference fields that the network is trained to predict. (In W22, they were called ``truth" fields.) In \ch{computations}, equations \eqref{eq:eularian filtering1} -- \eqref{eq:eularian filtering2} are evaluated at each $(x,y)$ by harmonic fitting (least-squares method) from finely sampled time series with time intervals of $P/144$ \ch{over fitting windows of 1 day ($2P$) \cite{2015_GRL_PonteKlein}}. \ch{The fine} temporal \ch{sampling makes the} harmonic fitting \ch{over the short time window possible}.
\ch{Accordingly, the reference fields represent a locally coherent semidiurnal component estimated over a 1-day window. We nevertheless refer to these ITs as “incoherent,” following \citeA{2015_GRL_PonteKlein}, as the same IT field is incoherent on the longer, altimetry-relevant fitting windows of $O(10)$ days discussed in the Introduction. As}
explained in the Introduction\ch{, }this \ch{fine} temporal information is \ch{not available} in global satellite data, which motivates our snapshot-based approach. Our deep learning algorithm only has access to snapshots captured with time intervals of two days ($4P$); under such coarse temporal sampling, harmonic analysis would yield nearly zero fields in the mid-jet and down-jet regions in our datasets (Aur\'elien Ponte, private communication). In the training and testing of our deep learning algorithm, all snapshots are randomly reshuffled, which further ensures that no temporal evolution is accessible.  

We use simulation outputs under five different profiles of meridional density gradients. We refer to these five simulations as T1 to T5 in order of increasing density gradients. 
The variation of the density gradients affects the IT patterns in two ways. First, it adjusts the eddy kinetic energy budget and leads to five different levels of turbulent  BM activity (hereafter ``turbulence levels"). A larger density gradient leads to a higher turbulence level, which induces stronger scattering of ITs, making the IT signals more incoherent and the IT patterns more complex. This was established in \citeA{2015_GRL_PonteKlein} and quantified in detail in W22. 
Second, the different density profiles affect the mode-1 IT wavelengths as a consequence of the mode-wise dispersion relationships \cite{gerkema2008introduction}. From T1 to T5, the minimum of background mode-1 IT horizontal wavelength in the mid-jet regions decreases from 170 km to 140 km, as quantified in W22.

\subsection{Deep Learning Algorithm} \label{sec:dl_algorithm}
We cast the extraction of IT imprints as an image-to-image translation problem, and we apply convolutional neural networks to map \emph{inputs}, which are different combinations of unfiltered surface fields, to \emph{outputs}, which are the filtered IT signatures on SSH. In principle, we can also task our algorithm to learn IT signatures on other physical fields (e.g., on surface velocities); we stay focused on the IT in SSH as they are more relevant to SWOT's SSH observations.
W22 employed a cGAN. Here we instead use a U-Net, which is simpler, more efficient, and---when trained properly---achieves comparable skill. The U-Net has a symmetric encoder–decoder structure with skip connections. Although the U-Net is widely used as a baseline in the literature  \cite{azad2024medical}, W22 excluded it because early tests showed weak performance. 
We now understand that our previous failure with the U-Net was due to an inadequate design of the learning rate scheme in the training of U-Nets: the learning rate was kept constant, following the original work that first proposed the U-Net architecture \cite{ronneberger2015u}. In this work, we vary learning rates periodically with respect to number of epochs (number of iterations of datasets during training). 
The rationale is that a deep learning algorithm's training can be regarded as an optimization problem, with the loss function as the object to minimize. Small learning rates 
\ch{make smaller parameter updates and can help refine the solution more gradually, but they slow the training and may cause the optimization to settle prematurely into poorer solutions} \cite{smith2017cyclical, liu2022super}.
By periodically amplifying the learning rates, the  
\ch{optimization can move away from such poorer solutions, while the subsequent return to small learning rates allows further refinement \cite{loshchilov2016sgdr}}.
\ch{Empirically, the periodically varying learning rate leads to much better performance than constant-learning-rate training in our experiments (Supporting Information Text S1 and Figure S1), allowing our U-Net to perform similarly to the cGAN;}
subtle differences are discussed in \S \ref{sec:Unet:discussion}. The main results are run on a NVIDIA A100-SXM4-40GB GPU and take about 25 minutes to train the U-Net for 300 epochs\ch{, and the skill [mean $R^2$, to be defined in \S \ref{sec:capturedmissed}, expression \eqref{R2}] on the held-out testing data  (to be described in \S \ref{subsec:inputoutput}) typically levels off at around 150 epochs.}
The low computational cost of the U-Net allows us to more freely explore different combinations of input and outputs. 

\begin{figure}
    \centering
    \includegraphics[width=0.99\linewidth]{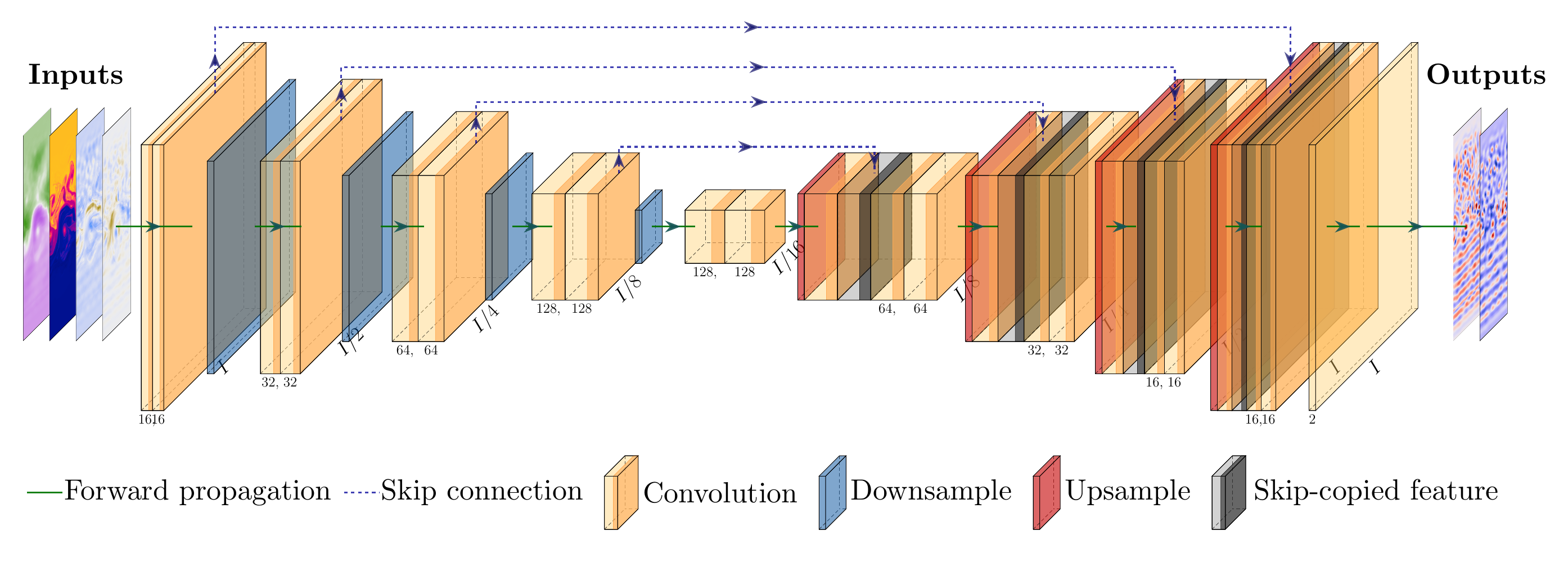}
    \caption{Schematic of the U-Net architecture employed in this work. The network takes combinations of surface fields in ($H$, $\boldsymbol{U}$, $T$) as input and predicts two output channels corresponding to the IT imprints on SSH, $h^{\text{sim}}_{\cos}$ and $h^{\text{sim}}_{\sin}$.
    Labels below each block denote (a) the number of feature maps, i.e., channels (e.g., $16, 16$) and (b) the dimension along one direction of each feature map (e.g., $I, I/2$). 
    Example input and output snapshots are taken from the same snapshot as  in Fig.~\ref{fig:Bouss_setup}.  Plotting code is adapted from \citeA{haris_iqbal_2018_2526396}.}
    \label{fig:Unet architecture}
\end{figure}

Fig.~\ref{fig:Unet architecture} sketches the U-Net architecture we apply.  Each input physical field is a separate input channel, and the U-Net's topology remains the same for different combinations of input physical fields. The training loss function is the standard point-wise \ch{$L_1$}. More implementation details and design considerations are provided in Supporting Information Text S1 and in the production codes\ch{ of \citeA{han_wang_2026_19829961}, which include a notebook illustrating the architectural details, data pipeline, and training strategy.}

For each combination of input data, we re-initialize the U-Net with random kernel values and retrain it ten times. The statistics recorded in Supplemental Information Text S2 show that for each Configuration of input data discussed in this work, the ten retrained U-Nets show negligible variation.
In the main text, for simplicity, under each configuration of input data, we \ch{make the arbitrary choice of picking the first (in terms of computational clock) trained} out of the ten retrained U-Nets, and only report on the outputs from that one U-Net. 

\chm{Other than the random initialization of kernel values, our U-Net does not contain additional stochastic components (e.g., no dropout is used). The retraining over different random initializations is used here as a robustness check on the stability of the learned mapping under a fixed architecture and training setup. The spread across retrained U-Nets therefore reflects only sensitivity to initialization, and probes only a limited part of uncertainties associated with the model and training setup. In particular, it does not quantify uncertainty associated with alternative architectures, loss functions, hyperparameters, or training procedures. Moreover, our U-Net is trained to produce only point estimates of the target ITs, rather than parameters of a predictive distribution. 
Therefore, our trained U-Net is deterministic in the sense that it returns a single output for a given input, rather than a distribution of outputs.
Our experiments compare the skill of this deterministic U-Net across different combinations of input surface fields (specified in \S \ref{subsec:inputoutput}), and thereby quantify the relative usefulness of these inputs for this algorithm and benchmark dataset. They do not directly provide a decomposition of total error into contamination in the simulation-derived reference fields, incomplete constraint from the chosen inputs, and limitations of the model architecture or training procedure. Throughout, we interpret the error as potentially arising from all three sources.}

Like many other deep image-to-image translation algorithms, the U-Net does not apply a spatial filter in the same way as a transfer function. Instead, nonlinear activation layers enable it to learn complex, nonlinear mappings. A U-Net can exploit input fields (e.g., surface velocities) that do not contain the outputs (i.e., IT imprints on SSH) directly. This is the foundation for our explorations on the impacts of different input fields.


\subsection{Inputs and outputs} \label{subsec:inputoutput}
We now describe the inputs and outputs of our U-Net. To clarify terminologies, we defined \textit{inputs} and \textit{outputs} in section \ref{sec:dl_algorithm}, and sketched them in Fig.~\ref{fig:Unet architecture}. 
Both \textit{training} and \textit{testing} data consist of pairs of inputs and outputs. During training, the U-Net is exposed to pairs of inputs and outputs from the training data. After training, the U-Net is given inputs from the testing data and generates new outputs; the generated outputs are compared with the reference outputs from the testing data when we evaluate the performances. 

As introduced in \S \ref{sec:BoussSim}, our simulations are run at five different turbulence levels. W22 discussed different combinations of turbulence levels for the training and testing data. For example, in what they referred to as the ``ET5 run", the deep learning algorithm was trained on turbulence levels T1-- T4, and tested on T5. Similarly, ``ET1 run" uses T1 as the testing data. The training and testing data contained disjoint turbulence levels, which ensured that the ITs have different levels of incoherence and different characteristic spatial scales between the training and testing data, and tested the deep learning algorithm's ability to generalize. In W22, among the five runs ET1--ET5, the ET5 run was identified as performing the worst, both quantitatively and qualitatively. The same behavior is found in experiments with our U-Net (not shown). The poorer performance in the ET5 run is not surprising, as the ITs and the BMs are more complex in the testing data. To stay focused on the most challenging case, hereafter we exclusively discuss  ET5. 

In the Boussinesq model, after the ITs are forced, there is a ramp-up period of around 100 days where the ITs propagate and fill up the simulation domain. W22 excluded these snapshots while here we include them. We also include the wave-maker and sponge regions that were excluded in W22. These choices introduce a slightly greater diversity of dynamic regimes present in the data. We include 150 snapshots with time intervals of two days from each turbulence level. This results in 600 snapshots from T1–T4 in the training data  and 150 snapshots from T5 in the \ch{held-out} testing data \ch{(also referred to as ``testing data" in this work)}. 
\ch{The T5 data are not used for gradient-based parameter updates, but they are used to select the saved checkpoint across epochs and to report the performance metrics shown below.}

\begin{figure}
    \centering
    \includegraphics[width=0.99\textwidth]{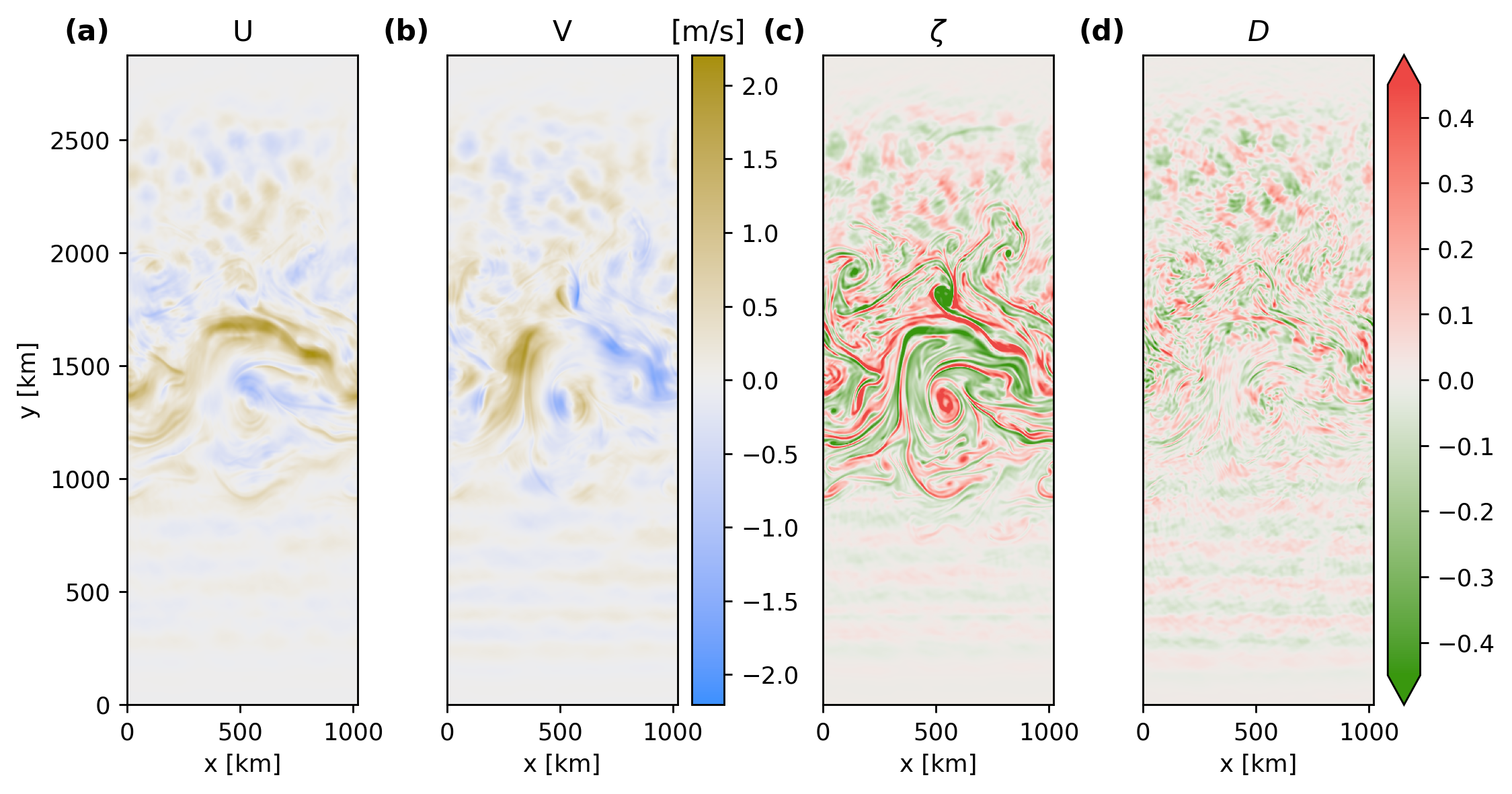}
    \caption{Surface zonal and meridional velocities $(U,V)$ (panels (a,b)), derived surface vorticity $\zeta$, and divergence $D$ (panels (c,d)), taken at the same snapshot  as in Fig.~\ref{fig:Bouss_setup}. The fields $\zeta$ and $D$ are computed by finite differencing the velocity field, and are non-dimensionalized by the Coriolis frequency evaluated at the central latitude of the domain (45 $^{\circ}$ N).
    }
    \label{fig: vel and vort inputs}
\end{figure}

The inputs of the model are different combinations of surface fields: the total SSH $H$, the surface temperature $T$, and the surface velocity fields $\bs {U} = (U, V)$, where $U$ and $V$ denote zonal and meridional velocities. These are fields represented in the Boussinesq simulations. A snapshot is shown in Fig.~\ref{fig:Bouss_setup} (panels (a,c)) and Fig.~\ref{fig: vel and vort inputs} (panels (a,b)). 
We exhaustively test all possible combinations of these three fields, first inputting each field in isolation, then every two-member combination, and finally all three fields together. 
We refer to each combination of input fields as a ``Configuration" followed by brackets specifying the input fields. For example, ``Configuration $\{H,\bs{U}\}$" refers to deep learning experiments that take $H$ and $\bs{U}$ as inputs. In W22, only Configuration $\{H\}$ was studied. 

When we use $\bs {U}$ as an input, we include both zonal and meridional velocities, $U$ and $V$, unless specified otherwise, as they are generally measured in tandem on ocean surface by the instruments (e.g., Doppler Scatterometers \cite{rodriguez2018optimal} and coastal high-frequency radars \cite{kachelein2024characterizing}) we are aware of. 

Under each configuration of inputs, the outputs that the U-Net tries to generate are  fixed to be $h^{\text{sim}}_{\cos}$ and $h^{\text{sim}}_{\sin}$ defined in \eqref{eq:eularian filtering1} -- \eqref{eq:eularian filtering2}, formatted as two output channels. They are the IT imprints on SSH, not on $\bs{U}$ or other input fields (see \S \ref{sec:dl_algorithm}, last paragraph, regarding why this works for  U-Nets).

\section{Impacts of different surface fields} \label{sec:impacts}
\subsection{Different inputs, different performances}
We now generally compare the U-Net performance under different inputs. 
We denote the ITs generated from our U-Nets (hereafter ``generated") as $h^{\text{gen}}_{\cos}$ and $h^{\text{gen}}_{\sin}$, which are the U-Net's attempted reconstructions of reference fields, $h^{\text{sim}}_{\cos}$ and $h^{\text{sim}}_{\sin}$, defined in \eqref{eq:eularian filtering1} -- \eqref{eq:eularian filtering2}; note the different superscripts, where ``gen" and ``sim" are shorthand for ``generated" and ``simulated" respectively. A snapshot in the mid-jet region generated from each Configuration compared against the reference are shown in Fig.~\ref{fig:Unet_tossh} for visual references. A visual inspection already suggests that some configurations perform better than others.

To quantify these impressions, here we evaluate two skill metrics between reference and generated fields:  
\begin{itemize}
    \item the correlation coefficient  $\Upsilon$, which rewards features in the generated images that vary linearly with the reference fields\ch{; its formal definition is
    \begin{equation} \label{upsilon}
    \Upsilon=\frac{\Sigma \left[(b - \langle{b}\rangle) (a - \langle a \rangle)\right]}{\sqrt{\Sigma (b - \langle{b}\rangle)^2\Sigma (a - \langle{a}\rangle)^2}},\end{equation}
    where $a$ and $b$ stand for the generated and reference fields respectively, the brackets or $\Sigma$ denote taking sample mean over or summing over all  grid points (and snapshots, if aggregated) considered,}
    and
    \item the coefficient of determination $\ch{R^2}$, which detects mismatches in overall magnitudes, or outliers where mismatches are large point-wise\ch{; its formal definition is
    \begin{equation} \label{R2}
    \ch{R^2} = 1- \langle{|b - a|^2}\rangle/\langle{|b - \overline b|^2}\rangle.
    \end{equation}}
\end{itemize}
The ranges for $\Upsilon$ and $\ch{R^2}$ are $[-1, 1]$ and $(-\infty,1]$ respectively, with values closer to $1$ indicating better alignment between the generated and reference fields.

At each $t$ in the testing data, we apply a trained U-Net to the input snapshot to obtain  $\left(h^{\text{gen}}_{\cos}(x,y,t), h^{\text{gen}}_{\sin}(x,y,t)\right)$, which is then flattened into a one-dimensional array. We do the same for $\left(h^{\text{sim}}_{\cos}(x,y,t), h^{\text{sim}}_{\sin}(x,y,t)\right)$. $\Upsilon$ and  $\ch{R^2}$ are computed between these two one-dimensional arrays at each $t$, and then averaged over all $t$.

\begin{figure}
    \centering
    \includegraphics[width=0.70\textwidth]{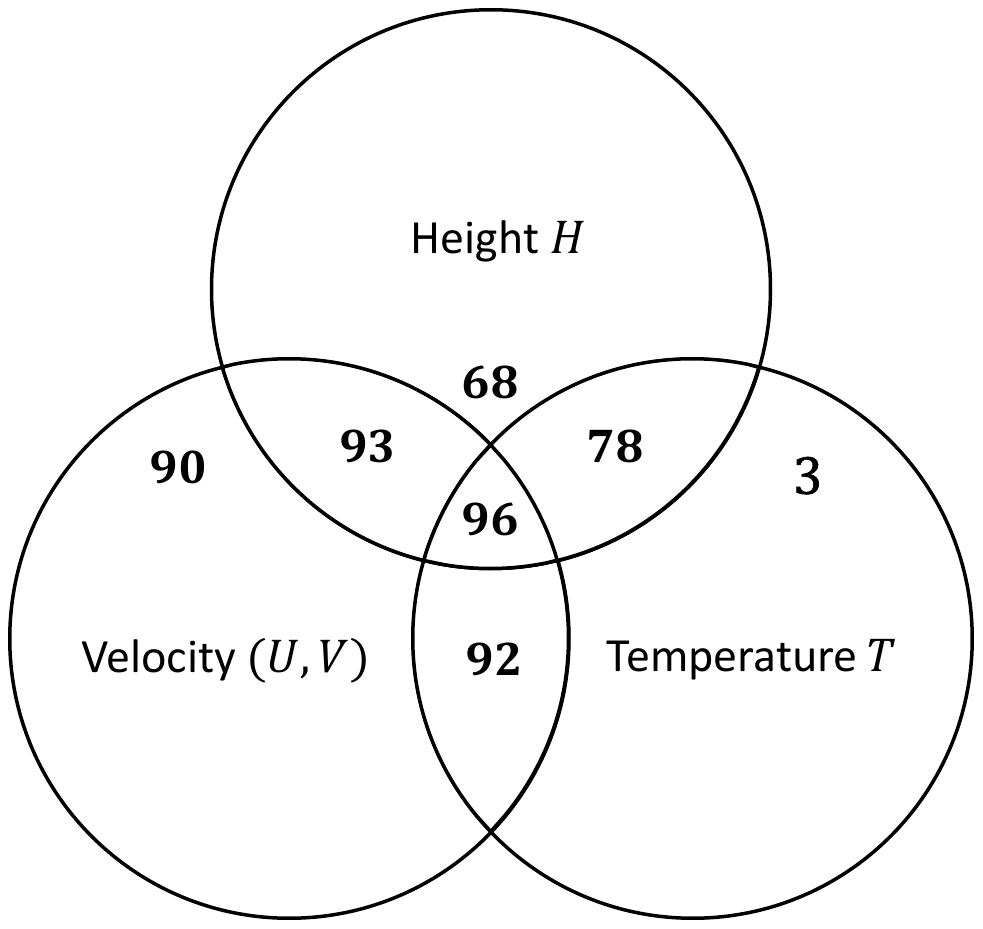}
    \caption{Summary of relative performances across input configurations. 
    Each circle corresponds to one input field in $(H,\bs{U},T)$; overlaps denote configurations that combine input fields. Numbers within circles show correlation ($100 \Upsilon$) averaged over the mid-jet region. 
    }
    \label{fig: UnetperformanceDiagram}
\end{figure}

\begin{figure}
    \centering
    \includegraphics[width=0.99\linewidth]{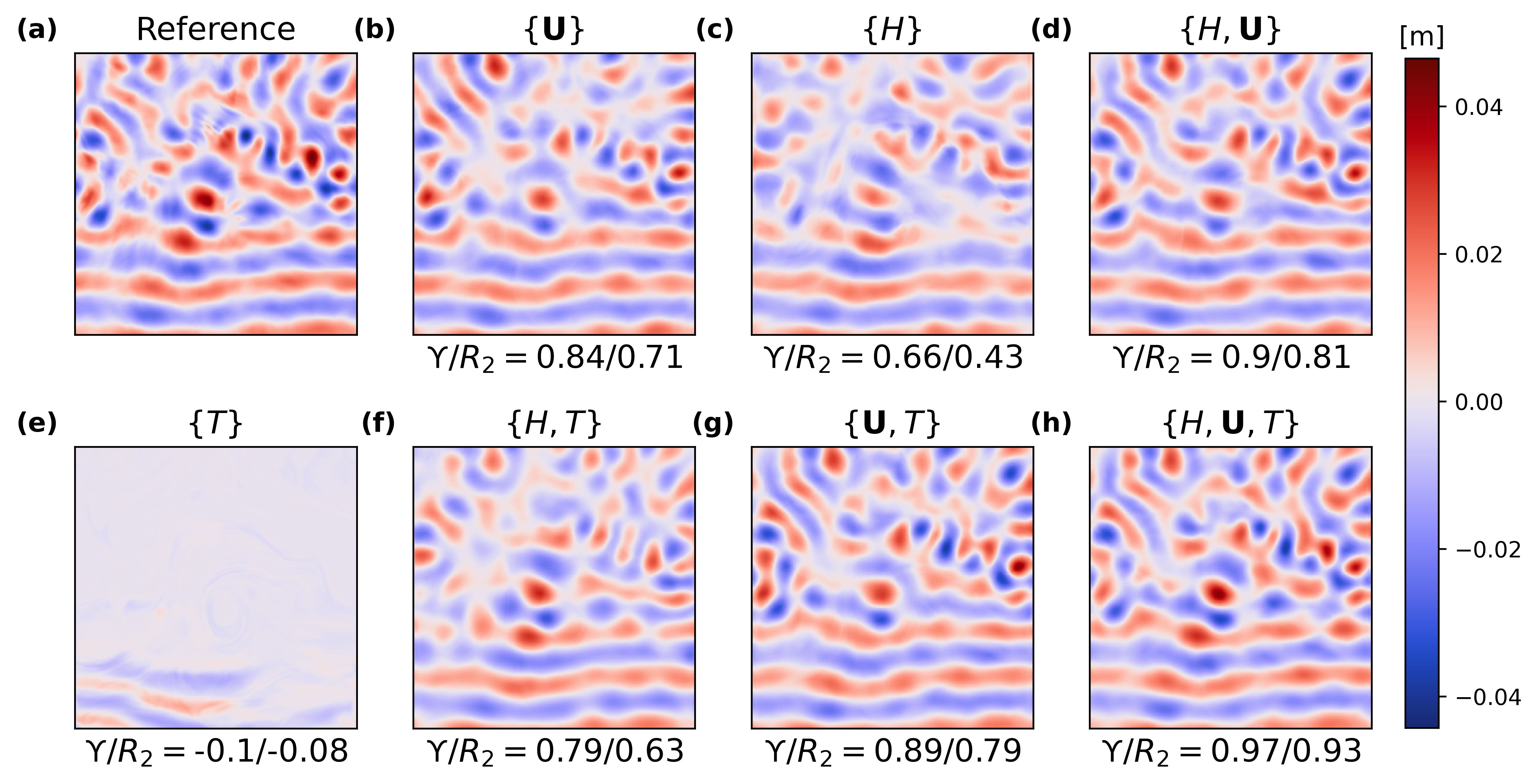}
    \caption{Comparison of reference IT signal $h^{\text{sim}}_{\cos}$ (panel (a)) with U-Net reconstructions $h^{\text{gen}}_{\cos}$ from different input configurations (panels (b-h)).  Numbers below each panel in (b-h) report $\Upsilon$ and $\ch{R^2}$ computed between the respective panels and panel (a). Only the mid-jet region is plotted, taken at the same snapshot as in Fig.~\ref{fig:Bouss_setup}. 
    }
    \label{fig:Unet_tossh}
\end{figure}

In addition to the $\Upsilon$ and $\ch{R^2}$ computed over all horizontal grid points in the simulated region, we also show  $\Upsilon$ and $\ch{R^2}$ computed over only the mid-jet regions. This is done by truncating $h^{\text{gen}}_{\cos}(x,y,t)$ and so on in $y$ leaving only the mid-jet regions, before the arrays are flattened. 
In W22, the mid-jet regions were identified to be more challenging than the other regions, hence the focus here. (In \S \ref{sec:capturedmissed}, we will elaborate more on the nature of challenges in the mid-jet regions.) 
Table \ref{tab:Unet ssh performance} shows $\Upsilon$ and $\ch{R^2}$ for all configurations for the full regions and the mid-jet regions. These metrics turn out to improve in tandem when we compare between configurations; for example, a configuration with higher $\Upsilon$ over the full region also has higher $\ch{R^2}$ over the mid-jet region. Therefore, we can identify which configuration performs better than another one with no ambiguities. Fig.~\ref{fig: UnetperformanceDiagram} is a concise summary on the relative performances between configurations; for brevity, only the mid-jet $\Upsilon$ is shown. 


\begin{table}
  \begin{center}
\def~{\hphantom{0}}
  \begin{tabular}{lcccccccc}
         Metric & $T$ &$ \bs {U}$ &  $H$ & $H, T$ & $\bs {U},  T $ & $H, \bs {U}$ & $H , \bs {U} , T$\\[3pt]
        \hline
       $\Upsilon \times 100$ &42/3.0 & 94/90 & 79/68 & 85/78 & 95/92 & 96/93 & 97/96\\
       $\ch{R^2}\times 100$ &16/-2.7 & 88/81 & 63/46 & 72/61 & 90/84 & 91/86 & 95/92
  \end{tabular}
  \caption{$\Upsilon$ and $\ch{R^2}$ for each Configuration. Values are reported as ``full / mid-jet.” The headings note the inputs in the Configurations.}
  \label{tab:Unet ssh performance}
  \end{center}
\end{table}

\subsection{Wave signature and scattering medium} \label{sec:wave_sig_scatter_med}

Clearly, having different input fields affect the performances. We discuss the \ch{physical} origins of these differences as follows. 
Each input field can contribute in two ways: (a) by exhibiting a measurable signature of the ITs, which aids the identification of the phase and orientation; and (b) by reflecting properties of the BMs and background conditions that modulate the ITs, which helps to interpret the IT energy distribution and overall statistics in a time-averaged, non-phase-resolving  sense. 
We refer to these two types of information as ``wave signature" and ``scattering medium".

ITs induce SSH fluctuations, and so conversely, $H$ contains wave signatures. It also provides information about the scattering medium. Part of the BMs that scatter ITs can be inferred from SSH. In our simulation data, the BM velocities are strongly correlated to geostrophic velocities proportional to the spatial gradients of SSH \cite{Ponte2017}. 
BM velocities that are not geostrophic also have a signature on SSH; for example, under the effective surface quasi-geostrophy model \cite{lapeyre2006dynamics} applicable to surface intensified eddies, SSH is related to geostrophic stream functions at all depths, which prescribes the entire BM dynamics. This is partly why deep learning approaches to extract BMs from SSH show promising performances \cite{xiao2023reconstruction,Gao2024}.
More speculatively, if single modes of ITs behave as shallow-water waves (an approximation roughly applicable to our simulation data \cite{LeGuillou2021}), the equivalent depth of each vertical mode is linked to SSH (\citeA{gill2016atmosphere}, p161), which dictates propagation metrics such as group speeds. 
For Configuration $\{H\}$, the main challenge of the U-Net is to disentangle the waves and the scattering medium (e.g., BMs), utilizing the differences and connections between the two. One particular challenge is that the imprints of ITs on SSH are about two \ch{orders of magnitude} smaller than the imprints from BMs in our simulation; in realistic ocean, internal wave imprints on SSH are also generally small compared to BMs, as a consequence of the large air-sea density differences. Thus, under this Configuration, our U-Net needs to disentangle signals that are very uneven in \ch{orders of magnitude}. 
This disentanglement is moderately successful, to the extent summarized in Fig.~\ref{fig: UnetperformanceDiagram}, and to be further detailed in \S \ref{sec:capturedmissed}. We next discuss the impacts of the other input surface fields through the same perspective. 

\subsection{Surface velocities}
From Fig.~\ref{fig: UnetperformanceDiagram}, Configuration $\{\bs{U}\}$ outperforms Configuration  $\{H\}$ or $\{T\}$ by a large margin. 
When $\bs{U}$ is added to other inputs, the improvements in performance are also significant.
We now discuss why using $\bU$ leads to such strong performances.

As mentioned before, we include both velocity components $(U,V)$ in the inputs because they are usually measured together at the ocean surface. 
One might still worry that Configuration $\{\boldsymbol{U}\}$ performs better than $\{H\}$ or $\{T\}$ simply because it supplies two input channels instead of one. 
\ch{We do not interpret the improvement in this way. 
Across input configurations, the overall U-Net topology is kept fixed; the only architectural adjustment is the change in the first convolution required to accommodate a different number of input channels. Therefore, we interpret the improved performance primarily as arising from the richer physical information in the inputs, rather than from increased architectural flexibility. To partly separate channel count from information content,}
we also trained networks with only a single velocity component as input, 
Configurations $\{U\}$ and $\{V\}$ (see Supporting Information Text~S3). In these cases, the number of input channels is identical to Configurations $\{H\}$ and $\{T\}$, but the mid-jet $\ch{R^2}$ values are $0.48$ for $\{U\}$ and $0.60$ for $\{V\}$, compared with $\ch{0.46}$ for $\{H\}$. Both single-component velocity configurations therefore outperform the SSH-only case despite having the same number of input channels. This indicates that the superior performance of Configuration $\{\boldsymbol{U}\}$ 
is not merely due to an advantage from providing more channels to the network.

There is an abundance of wave signatures and information \ch{about the} scattering medium in $\bU$. 
\ch{For a locally plane-wave IT component of a given vertical mode, the perturbations of $H$ and components of $\bU$   obey the standard linear polarization relations: 
they share the same horizontal wavevector, which sets the horizontal scale, 
and the linear wave solution determines  the relation between amplitudes and phase offsets, which, together with wavevectors, set the locations of wave crests. Although our ITs can not be simply modeled by such plane waves, this heuristic argument suggests that }
$\bU$ contains useful \ch{information about} wave signature. 
On the other hand, BM velocities \ch{induce} both local and far-reaching spatial variations of wave energies, as suggested by several works \cite{rainville2006propagation, Wagner2017,2021_JFM_SavvaKV,uncu2024wave}. Hence, the imprints from BMs on $\bU$ can provide important information on the scattering medium.

From this perspective, the information contained in $\bU$ is abundant and complementary to that in $H$. The exact reason why Configuration $\{\bU\}$ outperforms Configuration $\{H\}$ is not  clear. 
\ch{
In \S\ref{sec:degraded}, we conduct additional experiments in which the input SSVs are low-pass filtered in space, thereby smoothing out much of its fine-scale content. We find that adding the smoothed $\bU$ still substantially improves the performance,  even at output scales finer than those strongly retained in the smoothed SSVs. These results suggest that the larger-scale contextual information in $\bU$, associated with the BM-related scattering medium, is helpful. At the same time, the degraded performance at high wavenumbers when $\bU$ is smoothed indicates that the fine-scale wave-signature information in $\bU$ is also useful.}
Theoretically, under a Helmholtz decomposition of surface velocities, linear waves dominate over BMs in the divergent component (e.g., \citeA{buhler2014wave}), which explains a source of the strong wave signatures in $\bU$. We thus speculate, from the perspective of the Helmholtz decomposition, that an additional advantage of $\bU$ is a relatively clean conceptual pathway to disentangle the waves and BMs; this is discussed in \S \ref{sec:Unet:discussion}.

The performance of Configuration $\{H, \bs{U}\}$ is higher than both Configuration $\{\bs{U}\}$ and  Configuration $\{H\}$ by large margins. This reflects that the two variables provide complementary information, and adding one to the other improves the performance. 

\subsection{Surface Temperature} \label{sec:Tinput}
\ch{The surface temperature $T$ provides mostly only information on scattering medium. The imprint of the ITs on $T$ is much weaker than the BMs in our simulation, as demonstrated in \citeA{Ponte2017}. 
\citeA{Ponte2017} attributes the weak IT imprint to the scaling of horizontal advection terms. 
The weakness of internal wave signatures (including ITs) on surface temperature is demonstrated  in more realistic global circulation oceanic models \cite{TorresKlein2018} too. Only in (so far) rare observations can signs of imprints of ITs be found in SST}; a relatively recent perspective can be found in \citeA{farrar2007sea}.
By contrast, BMs leave strong imprints on $T$ and are dynamically linked to $T$. For example, under surface quasi-geostrophy frameworks \cite{lapeyre2006dynamics},  surface temperature (or density) anomalies determine a boundary condition for BM evolution.
In our simulation, the strong correlation between $T$ and potential vorticity at both the surface and the interior is verified in \citeA{Ponte2017}. 

As $T$ contains little wave signature, it is hard to infer the phases and directions of ITs from $T$ alone. 
A snapshot generated by Configuration $\{T\}$, compared against the reference (Fig.~\ref{fig:Unet_tossh}, panels (a,e)), shows that Configuration $\{T\}$ captures some structures near the up-jet region where ITs are relatively coherent (similar performance is found within the up-jet region, which is not shown), but misses the rest of the mid-jet region where the ITs are incoherent. 
As our snapshots are captured exactly every $4P$, the phases of coherent ITs are fixed between snapshots. For such ITs, phase information is less crucial, and the U-Net can  extract some coherent IT signals from  $T$ alone.
In contrast, strongly incoherent ITs have phases that vary strongly between snapshots in our dataset.
Meanwhile, as BMs evolve slowly, the input $\{T\}$ varies little between snapshots.
\chm{Thus,}
similar inputs lead to very different outputs\chm{; in other words, the conditional distribution of outputs given the input is broad. In this situation, our deterministic, single-output} U-Net responds by outputting weak, near-zero signals to minimize the training loss. 
Consequently, Configuration $\{T\}$ performs poorly under all metrics, especially in the mid-jet region.

This highlights that unsurprisingly, under a dearth of phase-resolving information in the inputs, the U-Net performs badly at resolving variable phases.
To further test this, we have run another configuration (not shown) using only the BM components of $H$ and $\bs U$, computed by top-hat averaging following \citeA{Ponte2017}, as inputs. This configuration also performs poorly.

Nevertheless, $T$ still provides useful information on the scattering medium. When combined with other inputs that contain abundant wave signatures, $T$  helps the U-Net  disentangle IT and BM imprints. Indeed, as seen in Fig.~\ref{fig: UnetperformanceDiagram}, adding  $T$ improves the performance of all configurations. Notably, Configuration  $\{H, \bs{U}, T\}$ outperforms  Configuration $\{H, \bs{U}\}$, showing that $T$ contains complementary information absent in $\bs U$ and $H$. 

\ch{These results are broadly consistent with findings in \citeA{Wang2025MultiScale}, which  reports that adding $T$ to $H$ yields a modest but consistent improvement in the skill on wave-BM disentanglement.}


In addition to SST, other surface fields with tracer-like behaviors may also provide complementary information on the scattering medium. In our simulations, the salinity is kept constant and no bio-geochemical activities are simulated; in more complex simulations, impacts of sea surface salinity or chlorophyll as additional inputs can be experimented on. These fields are in principle observable in the ocean; for instance,   NASA's Aquarius maps sea surface salinity, and the Plankton, Aerosol, Cloud, ocean Ecosystem (PACE) mission maps chlorophyll globally, partly overlapping with SWOT.
\section{Captured and missed features} \label{sec:capturedmissed}
Judging by $\Upsilon$ and $\ch{R^2}$ reported in \S \ref{sec:impacts}, Configuration $\{H, \bs{U}, T\}$ achieves very high skill, with mid-jet $\Upsilon$ and $\ch{R^2}$ close to $1$. Within our set of experiments, this represents the best performance obtained when the network is given all three surface state variables.  In this section, we examine more closely what Configuration $\{H, \bs{U}, T\}$ still misses\chm{, and how the residual errors should be interpreted.} From \S \ref{sec:impacts}, Configuration  $\{H\}$ performs worse, and we also \ch{discuss} what types of errors are particularly aggravated in Configuration $\{H\}$.

\begin{figure}
    \centering
    \includegraphics[width=0.99\linewidth]{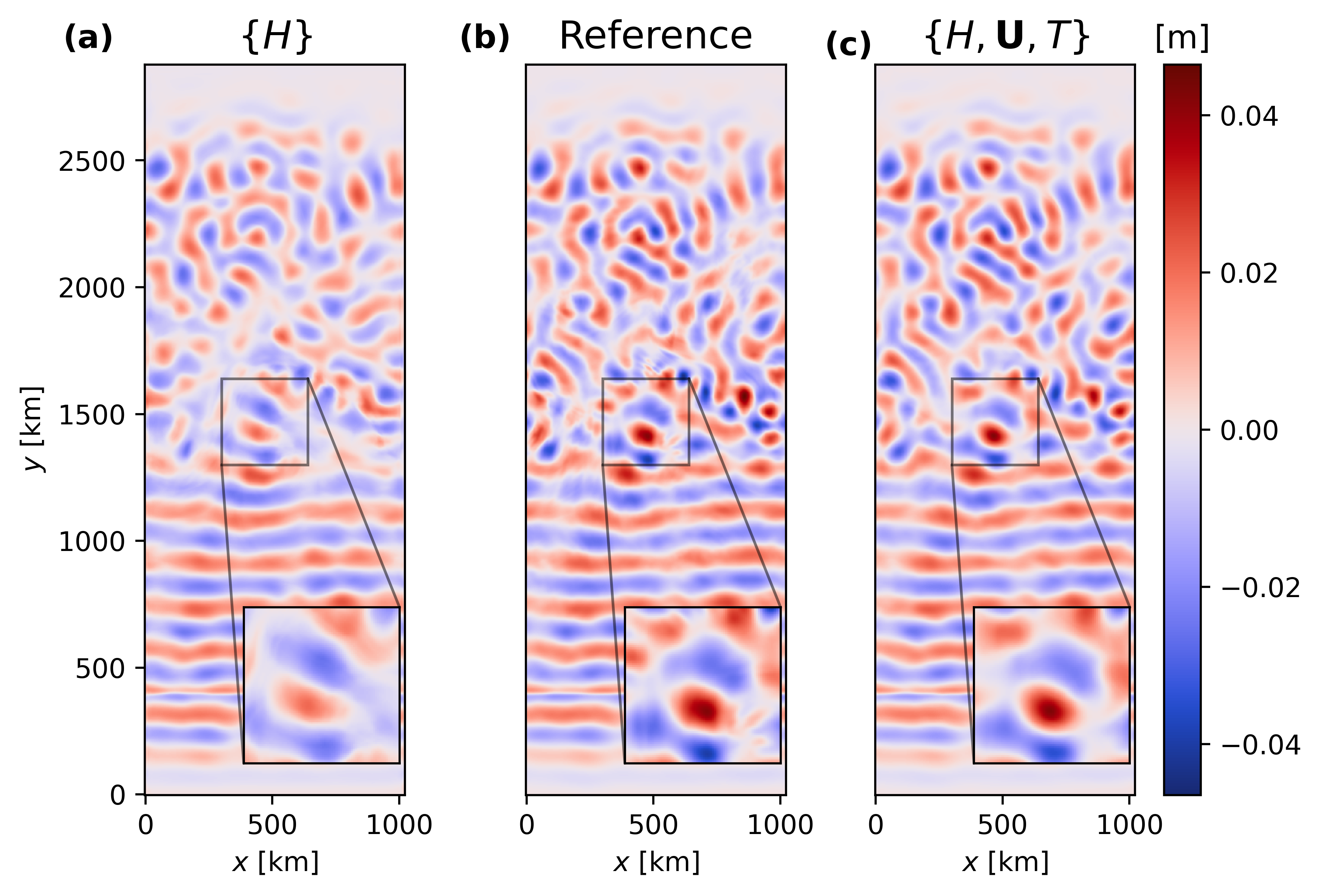}
    \caption{IT reconstructions $h^{\text{gen}}_{\cos}$ from Configuration $\{H\}$ (a) and Configuration $\{H, \bs{U}, T\}$ (c), compared to the reference $h^{\text{sim}}_{\cos}$ (b). Taken at the same snapshot as in Fig.~\ref{fig:Bouss_setup}.}
    \label{fig:Unet success and failure}
\end{figure}

We compute  the  $\Upsilon$ and $\ch{R^2}$ averaged over the up-jet/mid-jet/down-jet regions in the test data separately, and list them in Table \ref{tab:all panel performance}.
Performance in the up-jet is very high: even  Configuration $\{H\}$ scores well, and Configuration $\{H, \bs{U}, T\}$ improves further. 
This is consistent with the snapshots in Fig.~\ref{fig:Unet success and failure}, where visual agreements in the up-jet regions are excellent. 
As the ITs are more coherent in the up-jet regions, the IT patterns are more regular, posing fewer challenges to the U-Net. 

In contrast, the performances in the mid-jet and down-jet regions are significantly worse than in the up-jet, with the mid-jet region giving the poorest metrics. As noted in \S\ref{sec:BoussSim}, the mid-jet region has stronger meridional density gradients and more intense turbulent jet activity, which increase variations in background IT wavelengths and enhance scattering by BMs.
With the increased incoherence, the IT imprints \ch{(the target outputs)} are more complex in the mid-jet regions, 
posing more challenge to the U-Net.
This behavior was also demonstrated extensively in W22 for Configuration $\{H\}$, where a different neural network architecture (cGAN) was used; the increased challenge under stronger incoherence appears common between different neural networks. 

\begin{table}
  \begin{center}
\def~{\hphantom{0}}
  \begin{tabular}{lccc}
         Metric &  $\{H\}$ &  $\{H, \bs{U}, T\}$\\[3pt]
        \hline
       $\Upsilon \times 100$ & 79 (97/68/71)  & 97 (100/96/98) \\
       $\ch{R^2}  \times 100$ & 63 (94/46/50)  & 95 (99/92/96) 
  \end{tabular}
  \caption{$\Upsilon$ and $\ch{R^2}$ under Configurations $\{H\}$ and $\{H, \bs{U}, T\}$. The headings denote the Configurations. The statistics are presented in the format ``full (up-jet/mid-jet/down-jet)''.}
  \label{tab:all panel performance}
  \end{center}
\end{table}

\begin{figure}
    \centering
    \includegraphics[width=0.49\linewidth]{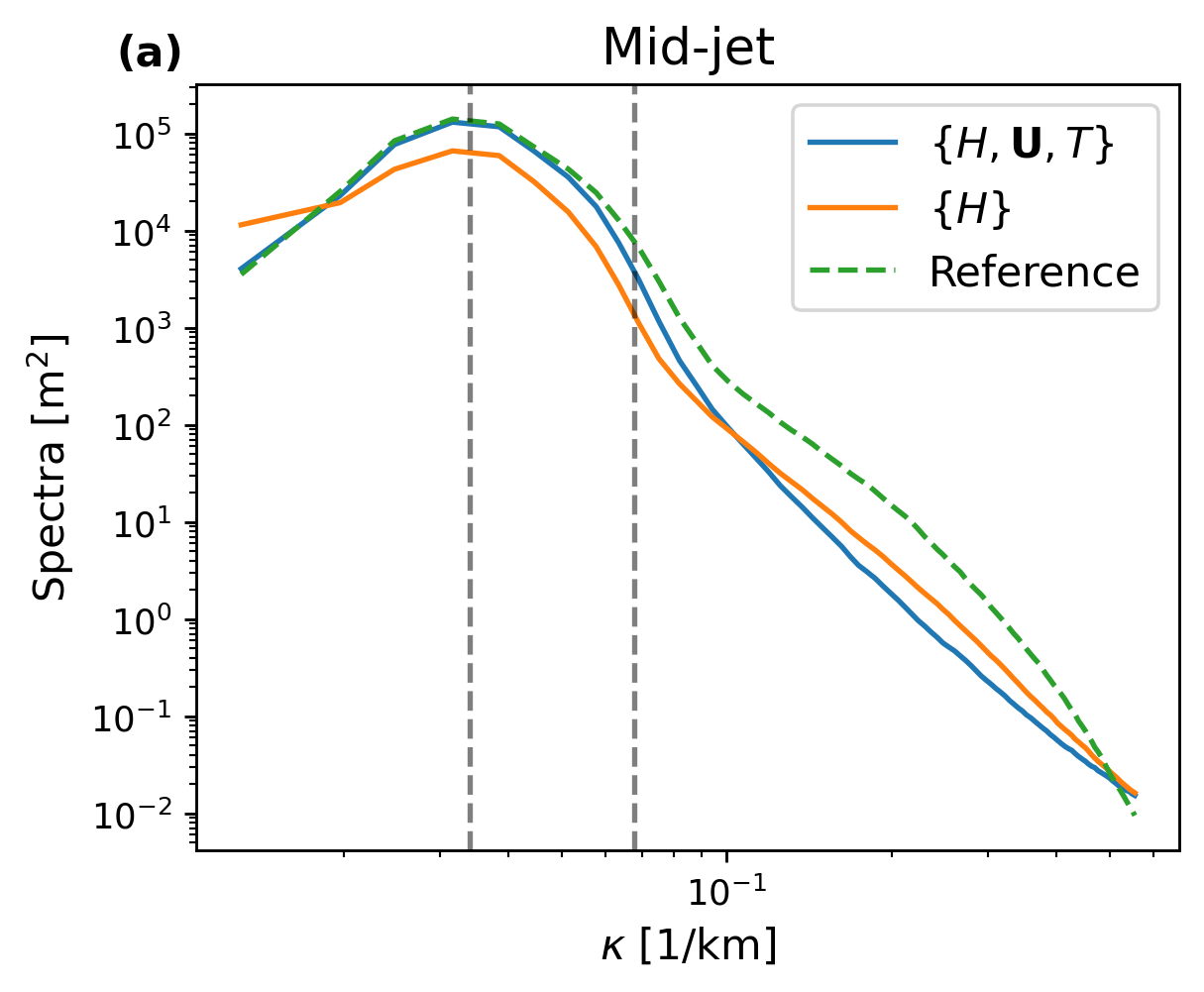}
        \includegraphics[width=0.49\linewidth]{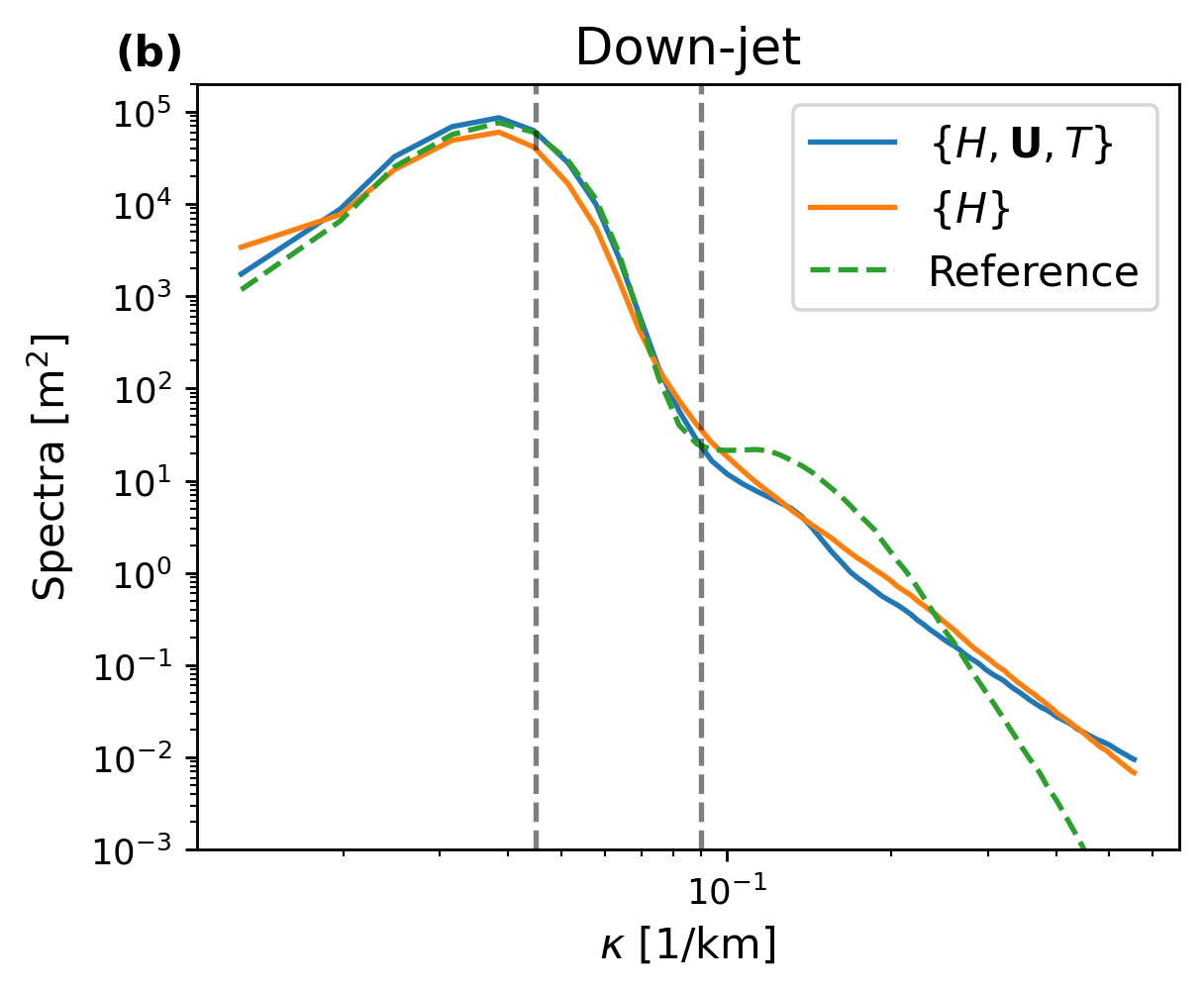}
    \caption{Azimuthally averaged wavenumber spectra of generated ITs $h^{\text{gen}}_{\cos}$ (input configurations denoted in the legends) and reference ITs $h^{\text{sim}}_{\cos}(x,y,t)$ (with legend label ``Reference"). Panels (a,b) correspond to the mid-jet and down-jet regions, respectively. \ch{The horizontal axis $\kappa\def\sqrt{k^2+l^2}$ is the radial distance of two-dimensional Fourier wave vectors.}
    Vertical dashed lines mark  mode-1 wavenumber and its \ch{second} harmonic ($2\times$), evaluated at the southern boundary of the mid-jet (panel (a)) or down-jet (panel (b)) regions. Spectra are averaged over the final 200 days (100 snapshots) of the testing set; values below $0.001$ m$^2$ or at $\kappa > 0.56$ km$^{-1}$ are omitted. 
    }
    \label{fig:Unet azim spectra}
\end{figure}

\begin{figure}
    \centering
    \includegraphics[width=0.49\linewidth]{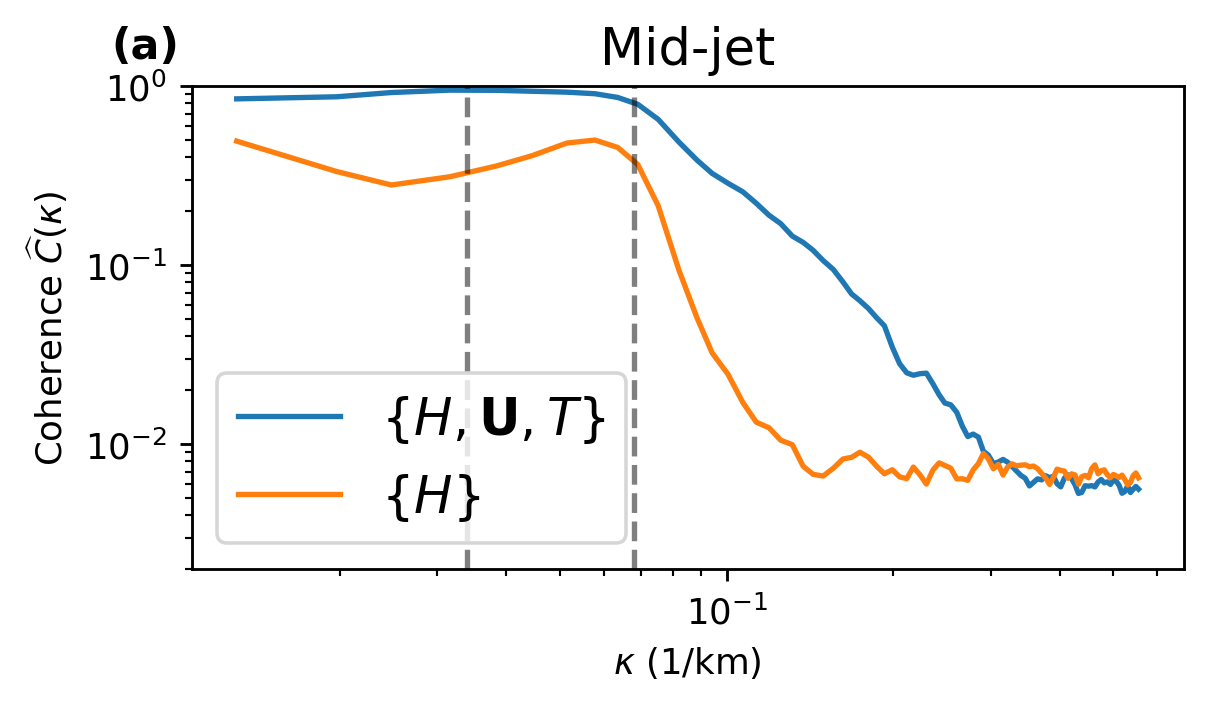}
        \includegraphics[width=0.49\linewidth]{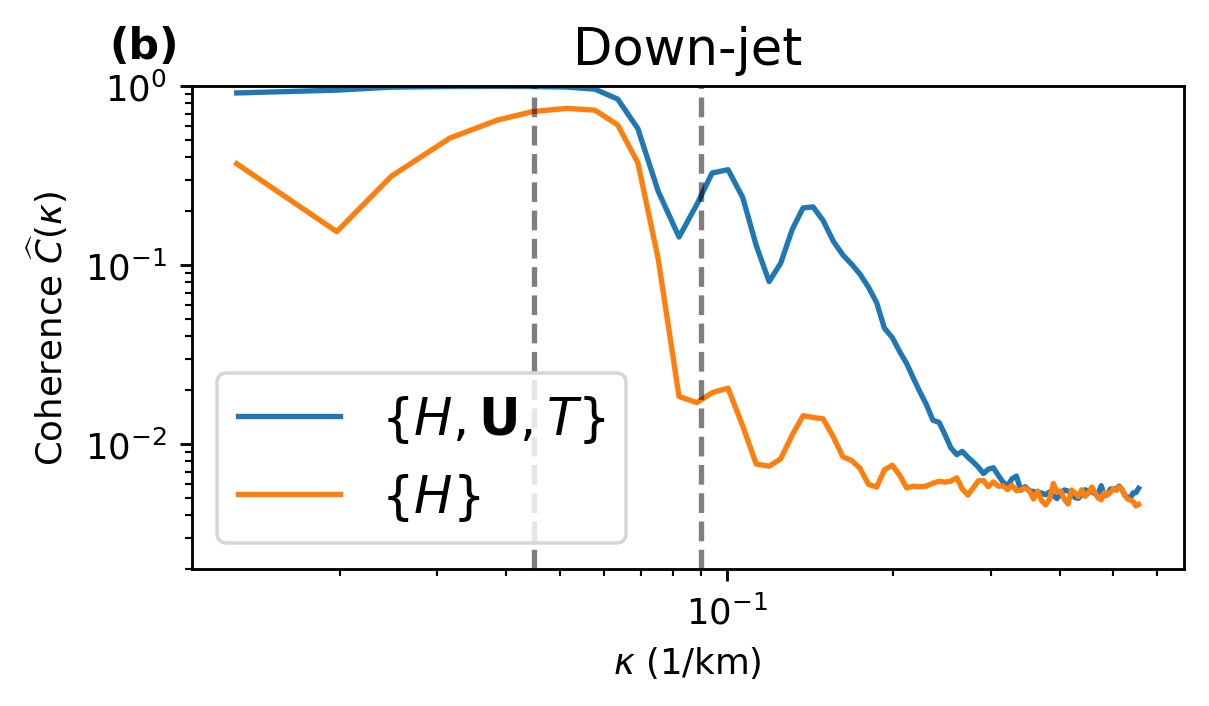}
    \caption{\ch{Azimuthal average of squared spectral coherence between the outputs from different configurations (marked in line legends) and the reference outputs.   
    Vertical dashed lines mark the incoming mode-1 wavenumber and its harmonic ($2\times$) as in Fig.~\ref{fig:Unet azim spectra}. The final 100 snapshots of the testing set are included in the sample mean; values at $\kappa > 0.56$ km $^{-1}$ are omitted.
    } }
    \label{fig:coh_HvsHUT}
\end{figure}

\ch{We also conduct a spectral analysis to assess the scale-dependent performance of our U-Nets. }
For brevity, we focus on \ch{averages of two-dimensional spectral metrics over the polar angles of horizontal wavenumber vectors  (hereafter ``azimuthally averaged" quantities). }
\ch{
Fig.~\ref{fig:coh_HvsHUT} shows $\coh(\kappa)$, the azimuthally averaged squared spectral coherence, defined as the azimuthal average of two-dimensional
squared spectral coherence:
\begin{equation}\label{coherence}
\coh(\kappa)=\frac{1}{2\pi}\int_{0}^{2\pi}\Tilde{C}(\kappa,\alpha)\,\mathrm{d}\alpha,
\end{equation}
where \((\kappa,\alpha)\) are polar coordinates in horizontal wavenumber space, with
\(k=\kappa\cos\alpha\) and \(l=\kappa\sin\alpha\), and
\begin{equation} \label{2Dcoh}
\Tilde{C}(\kappa,\alpha)=\Tilde{C}(k,l)=
\frac{\left|\left\langle \hat a^{*}(k,l)\hat b(k,l) \right\rangle\right|^2}
{\left\langle \hat a^{*}(k,l)\hat a(k,l)\right\rangle
\left\langle \hat b^{*}(k,l)\hat b(k,l)\right\rangle},
\end{equation}
where the hats denote Fourier coefficients, asterisks denote  complex conjugation, and the brackets denote sample mean over snapshots (whereas the brackets in \eqref{upsilon}-\eqref{R2} also imply an average over grid points). The quantity $\langle \hat a^{*}(k,l)\hat b(k,l)\rangle$ in the numerator of \eqref{2Dcoh} is the standard sample estimate of cross-spectrum, that is, the Fourier transform of cross-covariance \cite{yaglom1952introduction} between the reference and generated fields, $a$ and $b$. 
Fig.~\ref{fig:Unet azim spectra} shows the azimuthally averaged wavenumber power spectra (hereafter ``spectra") of ITs computed over mid-jet and down-jet regions, defined similarly as in W22. 
The spectra show the scale-dependent distribution of intensity of separate fields, while the coherence $\coh(\kappa)$ shows the scale-dependent distribution of the correlatedness between fields; they provide complementary information. }A Hanning window 
in both meridional and zonal directions is multiplied to each snapshot before the fast Fourier transforms are \ch{performed in the computation of spectra and $\coh(\kappa)$}. The windowing in the zonal direction is called for even though our Boussinesq simulation is zonally doubly periodic: our U-Net is not enforced to recognize the periodicity, and outputs from our U-Net are not guaranteed to be zonally periodic. 

In spectra from the reference fields, the locations of the peaks around mode-1 IT wavenumbers are different between the down-jet and mid-jet regions: the incoming mode-1 IT wavenumbers (the first vertical dashed line in each panel) are approximately $2\pi/185$ km$^{-1}$ and $2\pi/140$ km$^{-1}$ for mid-jet and down-jet. 
Around the incoming mode-1 IT wavenumbers, the spectra from Configuration $\{H, \bs{U}, T\}$ align well with the reference field. The accurate reconstruction of the varying locations of the peaks near mode-1 wavenumbers implies that, under Configuration $\{H, \bs{U}, T\}$, the U-Net recognizes  variations in the dominant spatial scales of ITs. 

The same cannot be stated for Configuration $\{H\}$, which shows tangible differences in the wavenumber spectra near the mode-1 peaks, especially in the mid-jet region. 
Descriptively, the errors of Configuration $\{H\}$ at spatial scales comparable to or larger than the mode-1 wavelengths can be divided into two types: 
\begin{enumerate}
    \item[(i)] distortions in the scattering patterns, i.e.\ incorrect spatial structure and orientation, and
    \item[(ii)] systematic underestimation of amplitudes.
\end{enumerate}

These two errors are visible in Fig.~\ref{fig:Unet success and failure} (comparing panels (a) and (b)), particularly in the zoomed-in regions at the center of turbulent jets. 
As errors of type (i) reflect disagreements in the trends of patterns, they affect  $\Upsilon$. 
As errors of type (ii) are mismatches of magnitudes, they affect $\ch{R^2}$.

Both errors (i) and (ii) are greatly alleviated in Configuration  $\{H,\bs{U},T\}$, as reflected in the example snapshot (Fig.~\ref{fig:Unet success and failure}), the metrics (Table \ref{tab:all panel performance}), the spectra (Fig.~\ref{fig:Unet azim spectra}) \ch{and the coherence $\coh(\kappa)$ (Fig.~\ref{fig:coh_HvsHUT})}. 
\chm{This suggests that, under Configuration $\{H\}$, the ITs are less tightly constrained by the available input than under Configuration $\{H,\bs{U},T\}$. In other words, part of the larger errors of types (i) and (ii) under Configuration $\{H\}$ likely reflects the more limited information available from SSH alone, rather than only limitations of the U-Net architecture or training. }

At wavenumbers larger than roughly twice the incoming IT wavenumber (the second vertical dashed line in each panel in Fig.~\ref{fig:Unet azim spectra} \ch{and Fig.~\ref{fig:coh_HvsHUT}}), even the maximally performing Configuration $\{H, \bs{U}, T\}$ exhibits much lower spectral energy than the reference\ch{, and $\coh(\kappa)$ much smaller than $1$.}
This \ch{is consistent with} Fig.~\ref{fig:Unet success and failure}, where generated ITs appear smoother than the reference. We describe this as a third type of error:
\begin{enumerate}
    \item[(iii)] blurring of small-scale patterns.
\end{enumerate}

Not all of this ``error” necessarily reflects physical IT signals being missed. 
The ITs are computed in the reference fields through frequency filtering (equations \eqref{eq:eularian filtering1}--\eqref{eq:eularian filtering2}), which assumes a separation in time scales between the ITs and BMs. 
However, the baroclinic jet in the BM contains energy near tidal frequencies, which is then artificially counted as ITs.  
Moreover, the frequency filtering \eqref{eq:eularian filtering1}--\eqref{eq:eularian filtering2} is based on time series at fixed Eulerian locations, and the BMs can Doppler-shift waves with other frequencies to the tidal frequency band. 
The simulation we use is designed so that waves are dominated by a single tidal frequency, so the wave energy that could be Doppler-shifted to the tidal frequencies is small in an averaged sense. Nevertheless, qualitative differences are reported in the behaviors of IT phases between Eulerian and Lagrangian recordings from simulations set up similarly to ours \cite{caspar2022characterization}. 
Finally, the computations of \eqref{eq:eularian filtering1} -- \eqref{eq:eularian filtering2} are realized by least-square fittings (detailed in \citeA{2015_GRL_PonteKlein, Ponte2017}), which can contain numerical inaccuracies. 
\chm{These mechanisms can contaminate the simulation-derived reference fields by introducing small-scale signals that do not belong physically as ITs. 
}

\ch{The possible contamination at small scales in the reference data implies that the detected small-scale blurring should not be interpreted completely as a sign of model failure. }
Nevertheless, part of the missed small-scale patterns corresponds to genuine ITs, and it's a failure of the U-Net to miss them. In particular, 
the secondary spectral peak near twice the incoming mode-1 wavenumber (seen in the down-jet region in Fig.~\ref{fig:Unet azim spectra}, panel (b)) arises from nonlinear scattering of mode-1 into mode-2. This physically meaningful signal is strongly damped in the U-Net reconstructions, even under
Configuration $\{H, \bs{U}, T\}$. Thus, error (iii) is partly a real limitation.

\chm{Some of the fine-scale features in the target outputs, whether physically meaningful ITs or spurious data introduced by contamination in the reference fields, may be only weakly constrained by a single $(H,\bs{U},T)$ snapshot. This may be especially relevant for small-scale ITs affected by strongly nonlinear processes such as wave--mean and wave--wave interactions. The distribution of plausible IT imprints given the instantaneous inputs may be broad.
In this situation, our deterministic single-output U-Net can reduce the training loss by favoring a conservative central-tendency reconstruction, thereby damping the small-scale components in the outputs. This interpretation is consistent with the strong smoothing behavior seen in Configuration $\{T\}$ (\S \ref{sec:Tinput}), where physical arguments suggest that the output is only weakly constrained by the input, and the output is strongly smoothed toward zero in most of the mid-jet region. This motivates probabilistic formulations as a natural direction for future work, as they could represent such uncertainty explicitly rather than collapsing the possible outcomes into a single smoothed prediction. We discuss this more in \S \ref{sec:Unet:discussion}.}

\begin{figure}
    \centering
    \includegraphics[width=0.9\textwidth]{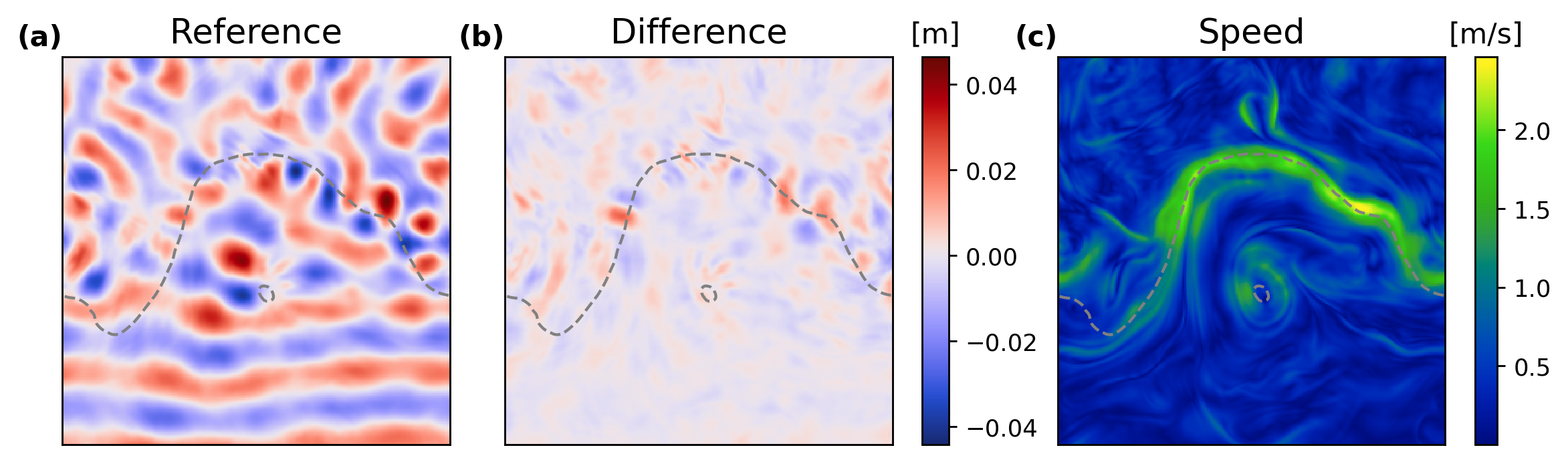}
    \caption{Panel (a): The reference $h^{\text{sim}}_{\cos}$. Panel (b): Difference field $\left(h^{\text{gen}}_{\cos}-h^{\text{sim}}_{\cos}\right)$ from Configuration $\{H,\bs{U},T\}$. Panels (a,b) share the same colorbar. Panel (c): Surface speed $|\bs{U}|$. Gray dashed contours show $H=0$ approximating the instantaneous jet axis. All are plotted over the mid-jet region at the same snapshot as in Fig.~\ref{fig:Bouss_setup}. }
    \label{fig:height jet}
\end{figure}

Fig.~\ref{fig:height jet} (panel (b)) shows a snapshot of the difference field $\left(h^{\text{gen}}_{\cos}-h^{\text{sim}}_{\cos}\right)$, where  $h^{\text{gen}}_{\cos}$ is generated under Configuration $\{H,\bs{U},T\}$.
This field is dominated by scales smaller than the mode-1 wavelength, consistent with error (iii). Interestingly, 
the regions with the strongest differences do not align with the regions of stronger reference ITs  (panel (a)), so the larger local errors are not due to stronger reference signals. Instead, the differences align with regions of large surface speed $|\bs{U}|$  (panel (c)).
Two mechanisms  can give rise to this. First, faster BM speeds and their gradients amplify spatially local wave-mean interactions \cite{2014_JGR_DunphyLamb,uncu2024wave}. 
\chm{In such dynamically complex regions, the distribution of plausible IT outputs given the  input snapshot may be broader than elsewhere, which would further favor conservative, smoothed reconstructions by the deterministic U-Net in these regions. This interpretation is broadly consistent with \citeA{Wang2025MultiScale}, which applies a probabilistic framework in a wave-BM disentanglement problem to find larger predictive uncertainty when BMs are dominant. While their uncertainty diagnostic is not identical to the local error field considered here, both results point to snapshot-based separation becoming harder where balanced motions are strong. 
}
Second, BMs near tidal frequencies contaminate the reference ITs that are computed via frequency filtering; these spurious contributions\ch{, which} are stronger where BMs are more energetic \cite{caspar2022characterization}, 
\chm{may be only weakly predictable from the input snapshot, and are therefore likely to be suppressed by the U-Net.}
For applications where behaviors of high-mode ITs are crucial, such as studies of energy dissipation pathways \cite{vic2019deep}, 
this systematic suppression of small-scale signals can be problematic. We discuss possible approaches to address this in \S \ref{sec:Unet:discussion}.

\begin{figure}
    \centering
    \includegraphics[width=0.8\linewidth]{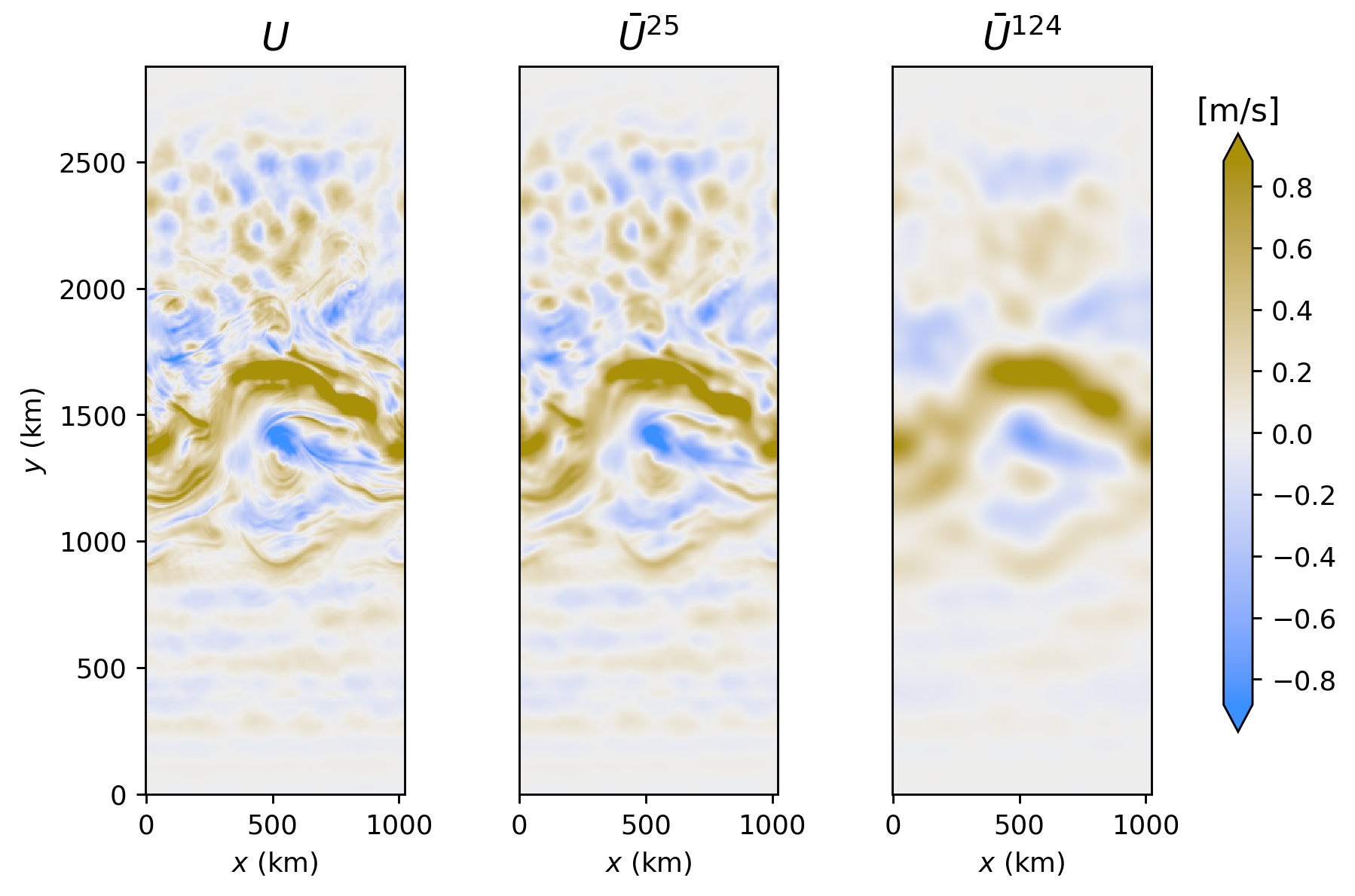}
    \includegraphics[width=0.8\linewidth]{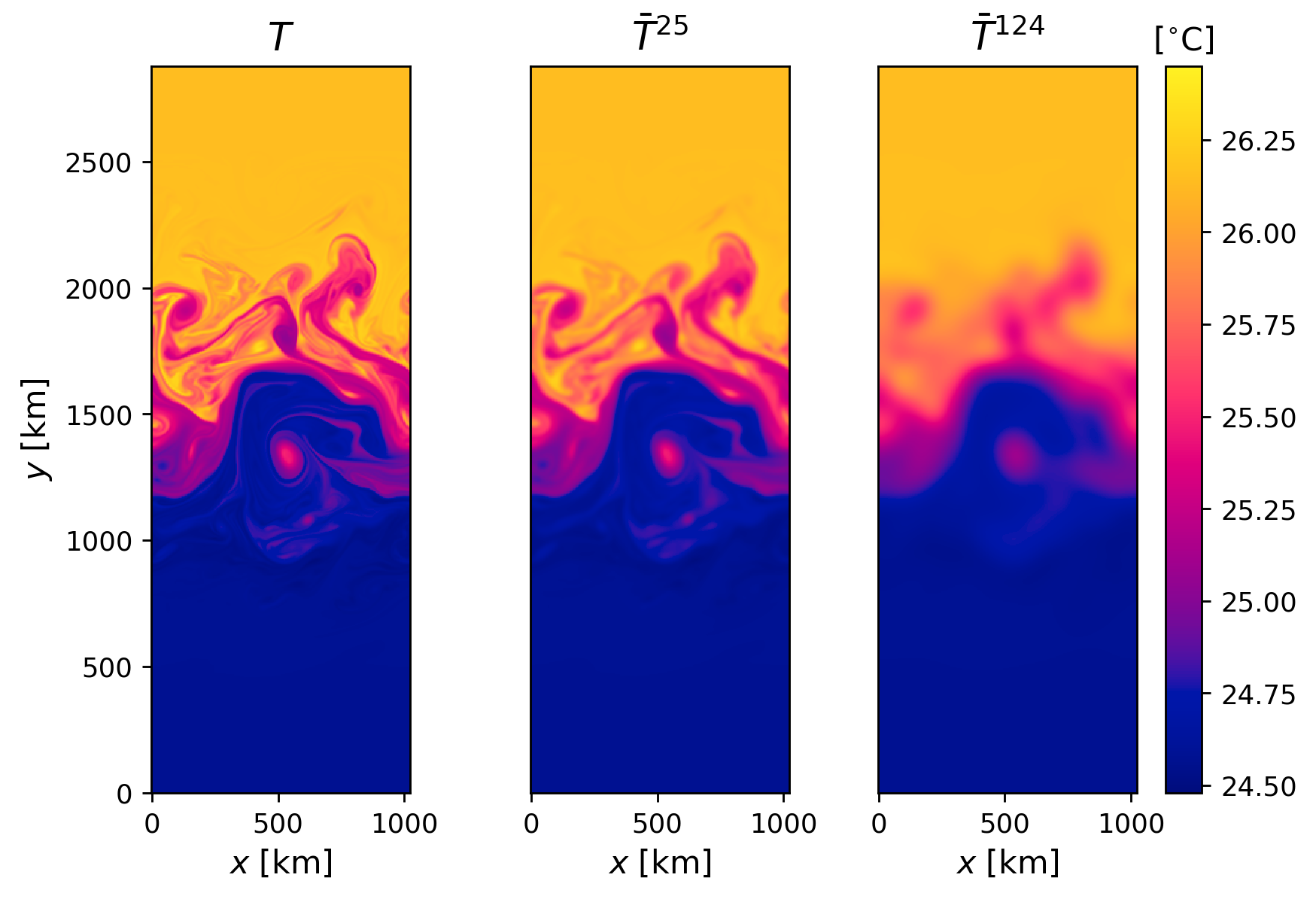}
\caption{\ch{Zonal velocity $U$ and surface temperature $T$ before and after lowpass filtering, at the same snapshot as Fig.~\ref{fig:Bouss_setup}.}}
\label{fig:UfiltandTfilt}
\end{figure}

\begin{figure}
    \raggedright
    \includegraphics[width=0.48\linewidth]{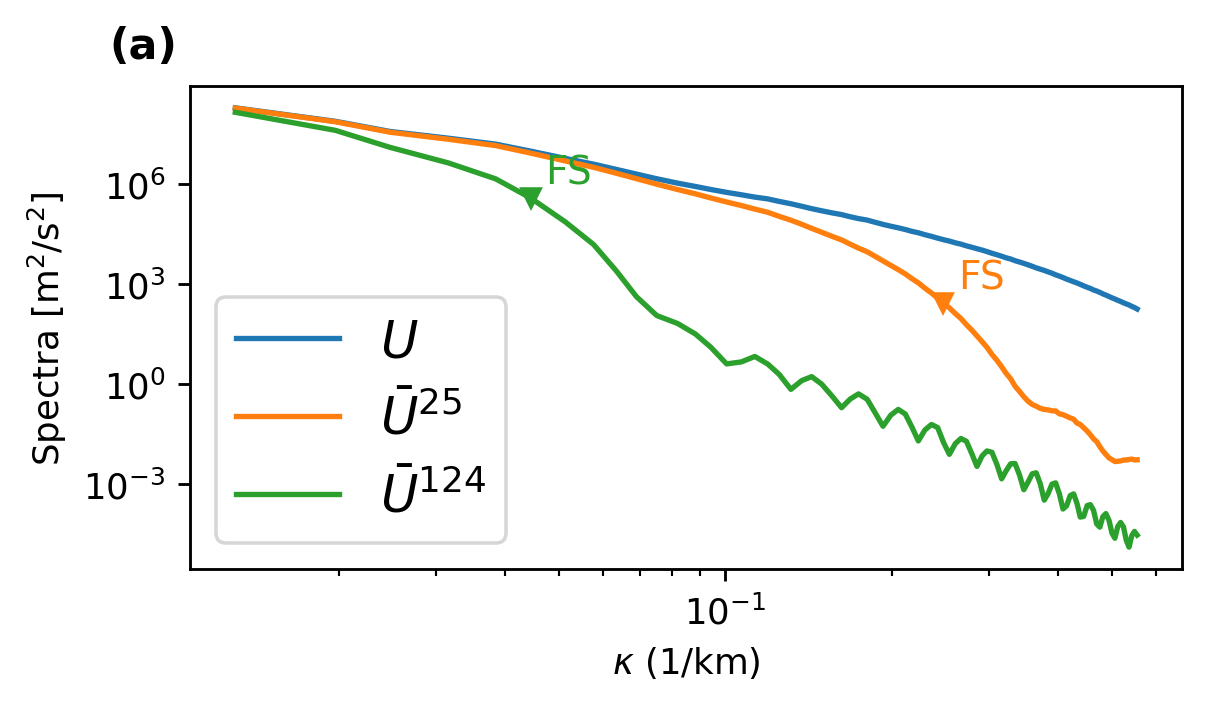}
    \includegraphics[width=0.48\linewidth]{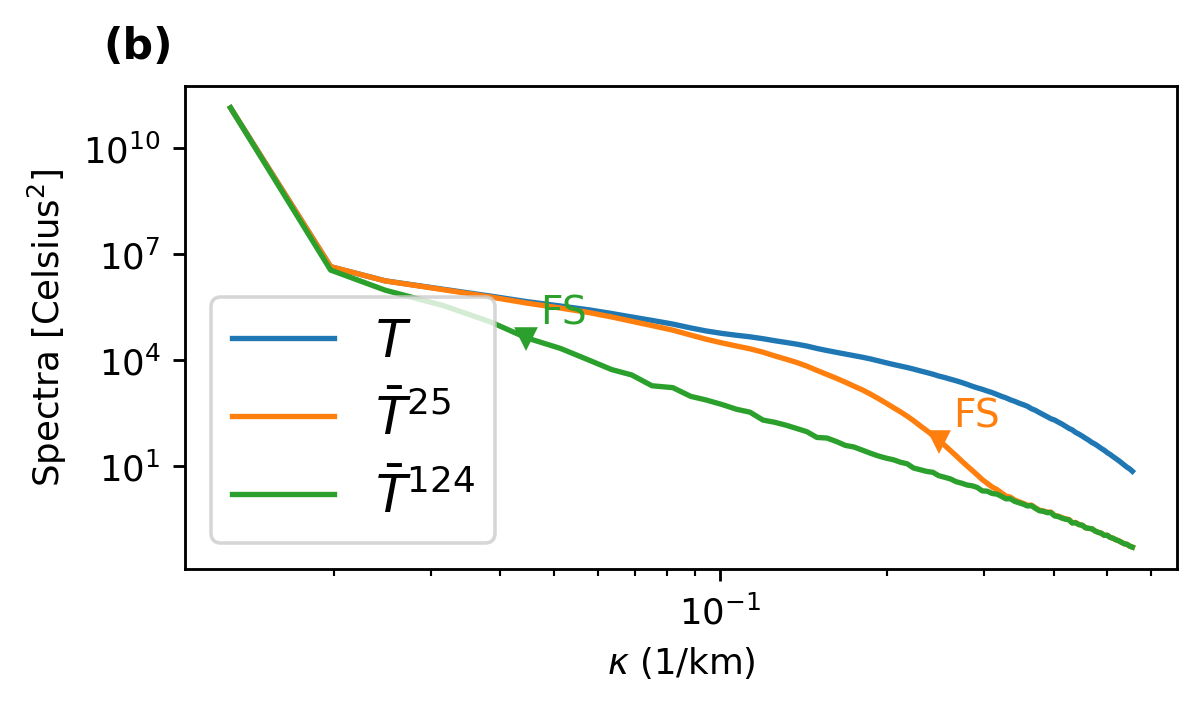}
    \includegraphics[width=0.48\linewidth]{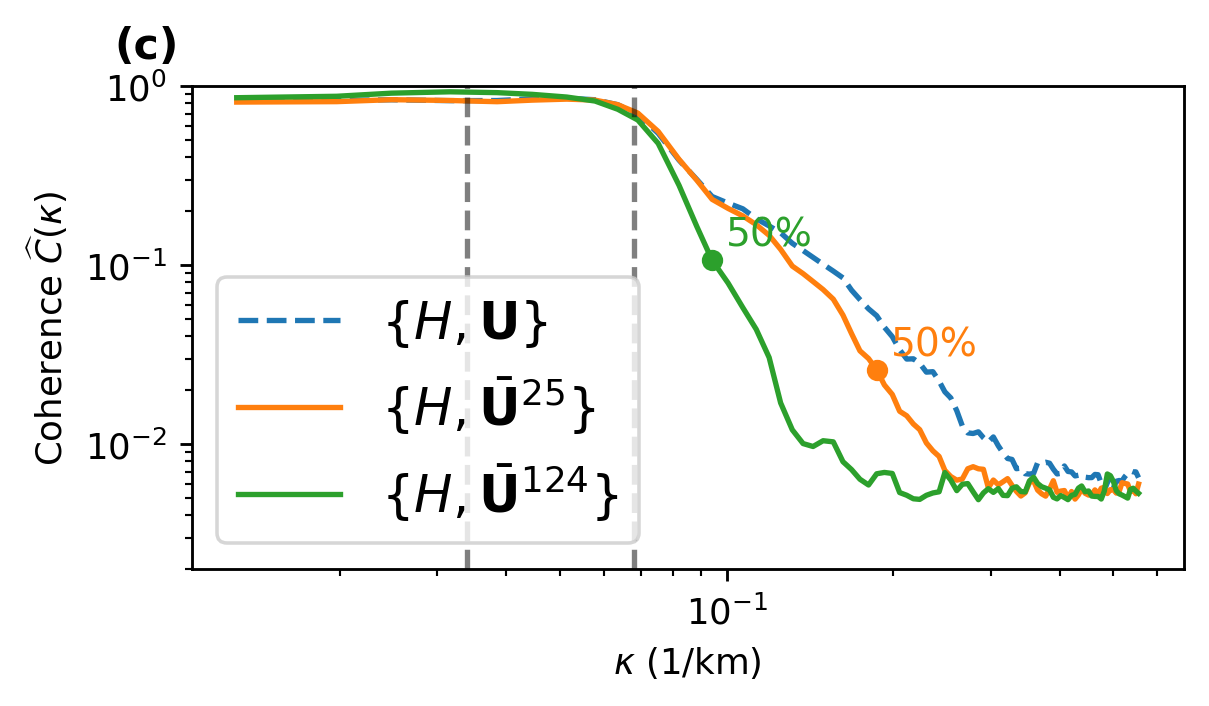}    \includegraphics[width=0.48\linewidth]{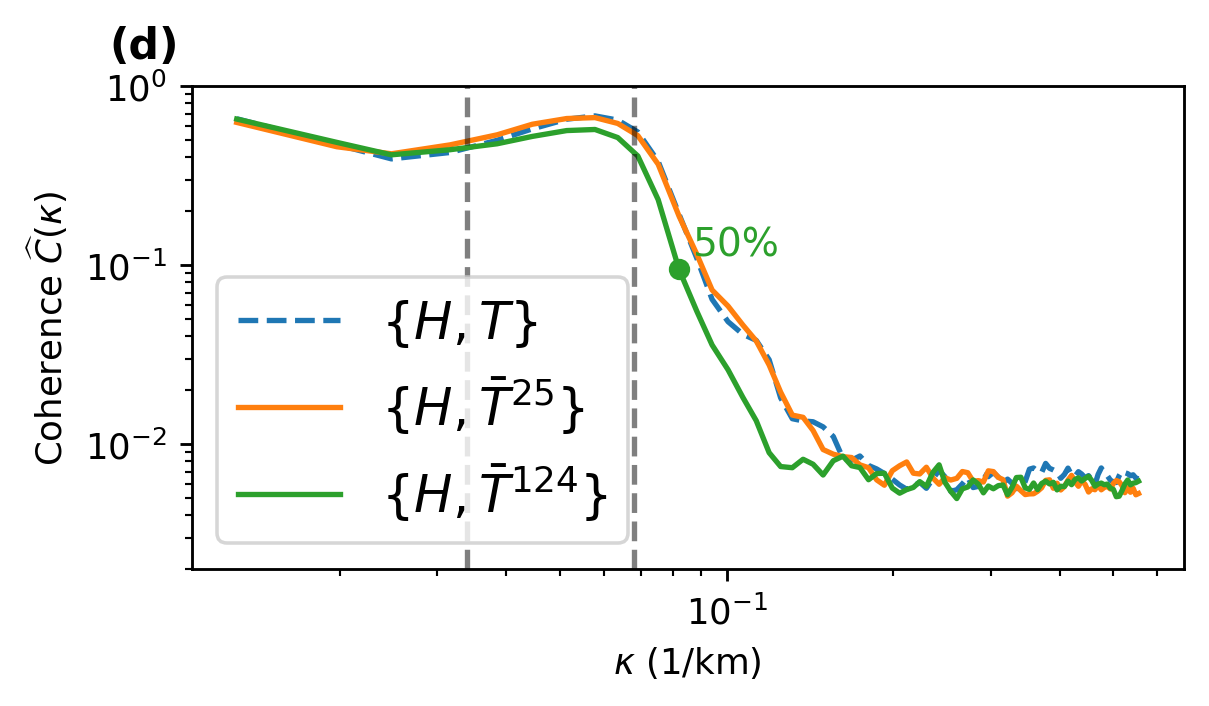}   
   \includegraphics[width=0.48\linewidth]{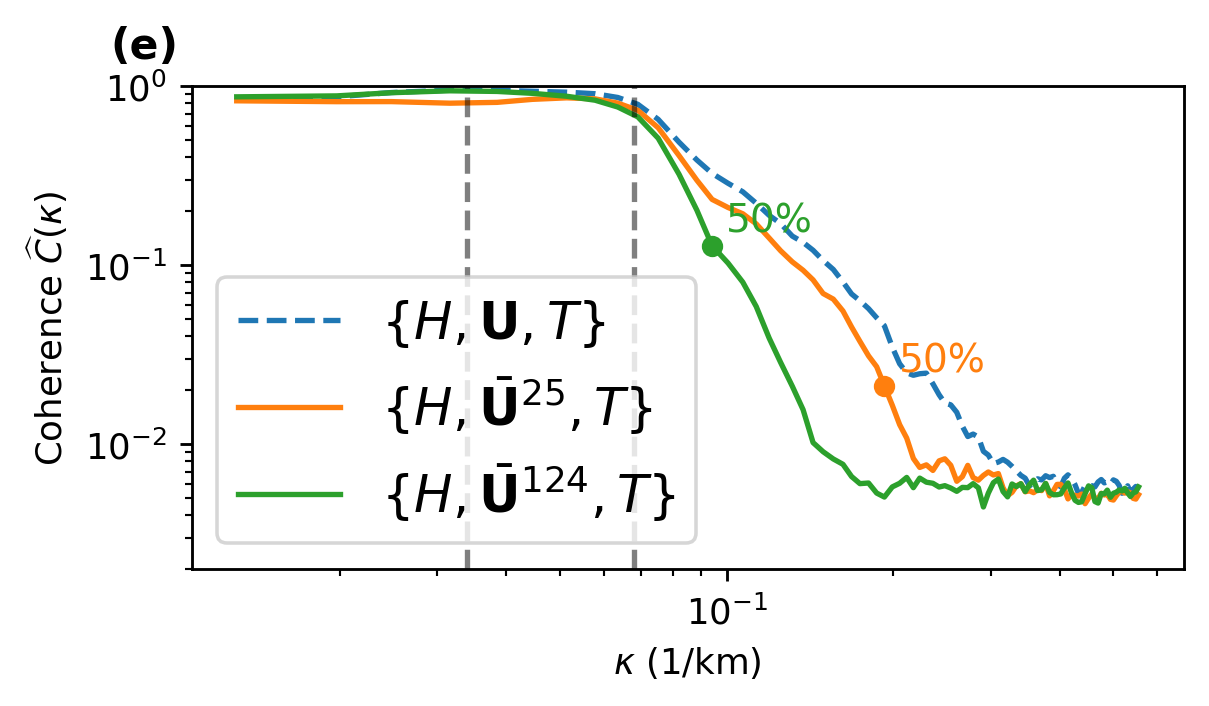}    \includegraphics[width=0.48\linewidth]{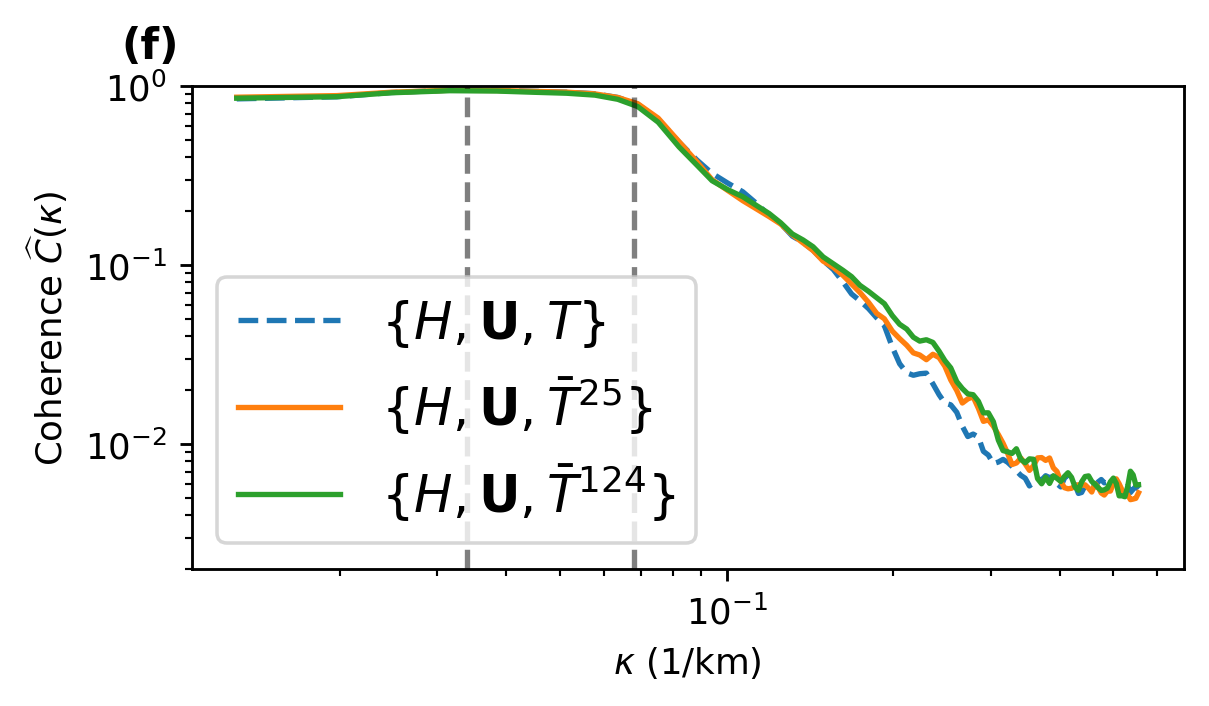}
    \includegraphics[width=0.48\linewidth]{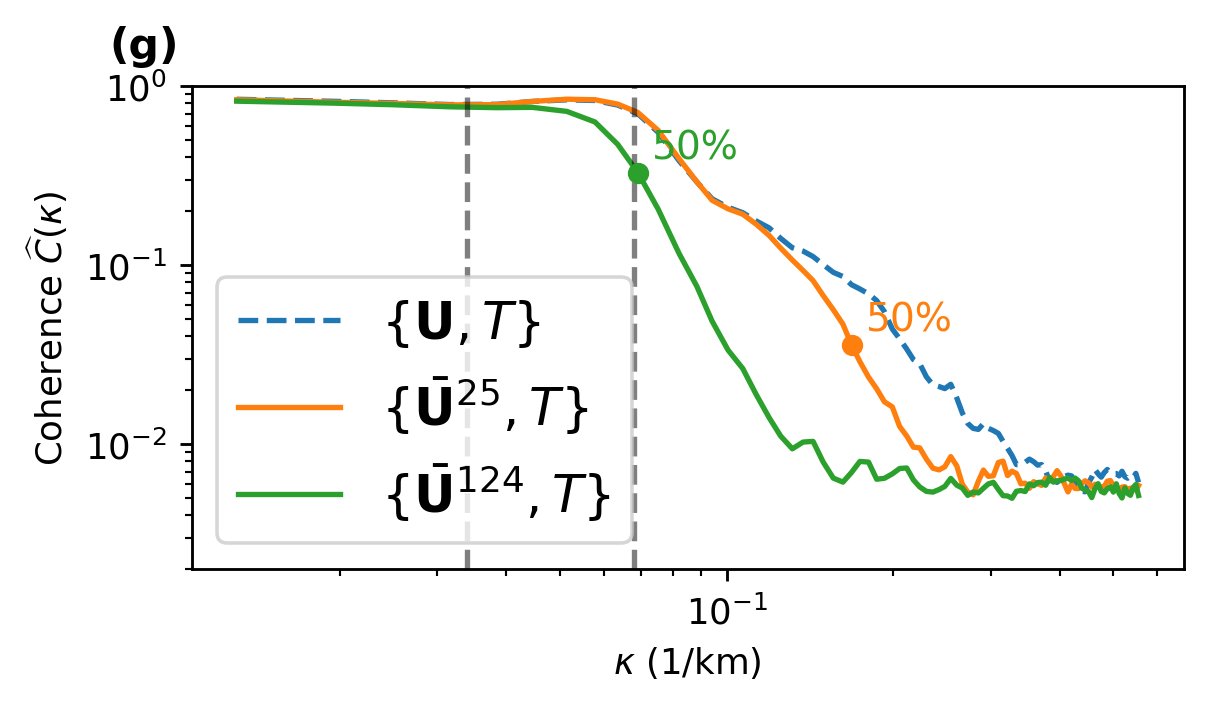}    \includegraphics[width=0.48\linewidth]{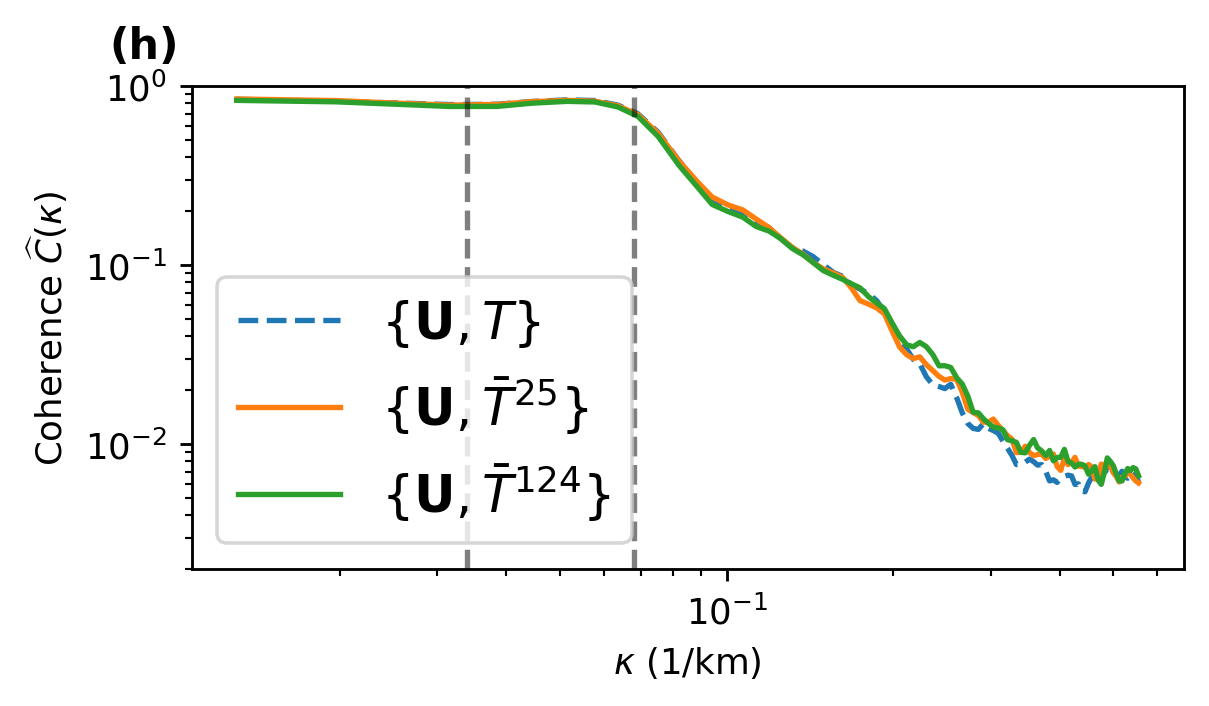}     
    \includegraphics[width=0.48\linewidth]{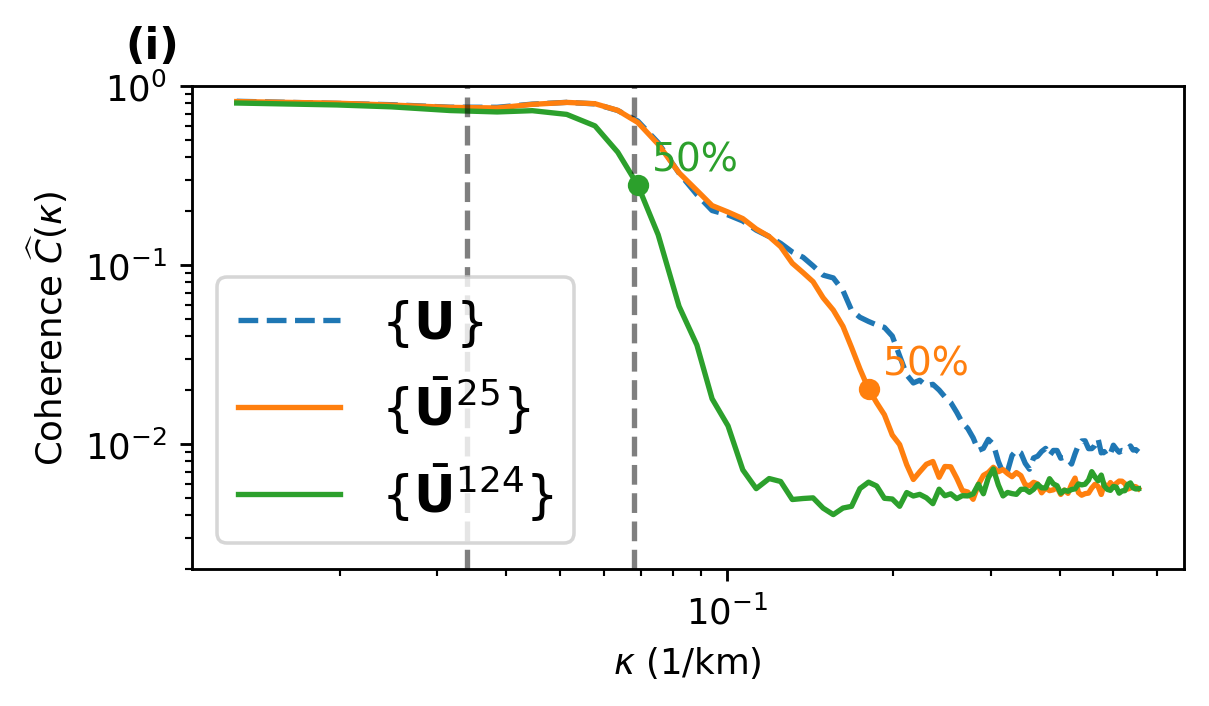}
\caption{\ch{Panels (a,b): spectra of $U$ or $T$ before and after smoothing. Triangles mark the filter scales of lowpass filters. Panels (c-i): coherence $\coh(\kappa)$ from configurations marked in the legend. The dashed curves are from unsmoothed-input configurations and solid curves are from smoothed-input configurations. Dots on the solid curves mark the smallest wavenumber where the ratio \eqref{coherenceratio} (i.e., the ratio between the solid curve and dashed curve) is ${}\leq 0.5$. 
Vertical dashed lines mark the incoming mode-1 wavenumber and its \ch{second} harmonic ($2\times$) as in Fig.~\ref{fig:Unet azim spectra}.
}}
\label{fig:coh_filteredconfigs}
\end{figure}

\ch{\section{Degraded inputs} \label{sec:degraded}
So far, in all our configurations with multiple inputs, the input fields are simultaneous (i.e., taken at the same snapshot) and represented on the native 4 km simulation grid. These conditions may not hold  in real observations. 
Motivated by practical limitations, here we probe two types of degradation: spatial smoothing of $\bs{U}$ or $T$, and temporal misalignment between input fields.
\subsection{Spatially smoothed inputs}
The horizontal grid spacing of our simulation is 4 km, which is comparable to the resolution of SWOT's processed SSH products [e.g., 2 km in \citeA{aviso_duacs_2024_swot_l3_lr_ssh}]. For SST, infrared retrievals under clear-sky conditions can approach a similar resolution, whereas cloud contamination, aerosols, and microwave-based retrievals generally lead to coarser effective resolution and/or larger gaps. For SSV, we are not aware of definite, quantitative resolution estimates for current satellite mission concepts; preliminary assessments for the HARMONY concept suggest that SSV could be reconstructed at a comparable spatial resolution \cite[Fig. 5]{kleinherenbrink2020ocean}. Motivated by the possibility that SST and SSV may in practice be available at coarser resolution than SSH, we apply Gaussian low-pass filters to $T$ and $\bs{U}$.
\\
We define a filter scale (hereafter ``FS'') as the wavelength at which the spectral transfer function of the Gaussian filter drops to $10\%$ of its peak value. The two filters used here have FS equal to $25$ km and $124$ km; the resulting filtered fields, say of $T$, are denoted by $\bar{T}^{25}$ and $\bar{T}^{124}$, respectively. A snapshot of $(U,\bar{U}^{25},\bar{U}^{124})$ and $(T,\bar{T}^{25},\bar{T}^{124})$ is shown in Fig.~\ref{fig:UfiltandTfilt}, where smoothing of the fine-scale features is visible in both variables. We also show the azimuthally averaged wavenumber spectra of $(U,\bar{U}^{25},\bar{U}^{124})$ and $(T,\bar{T}^{25},\bar{T}^{124})$ in Fig.~\ref{fig:coh_filteredconfigs}, panels (a,b), with the FS marked. The filtered fields differ more strongly from the original fields in $U$ than in $T$, as the unfiltered $U$ contains more variance at high wavenumbers.}

\begin{figure}
    \centering
    \includegraphics[width=0.48\linewidth]{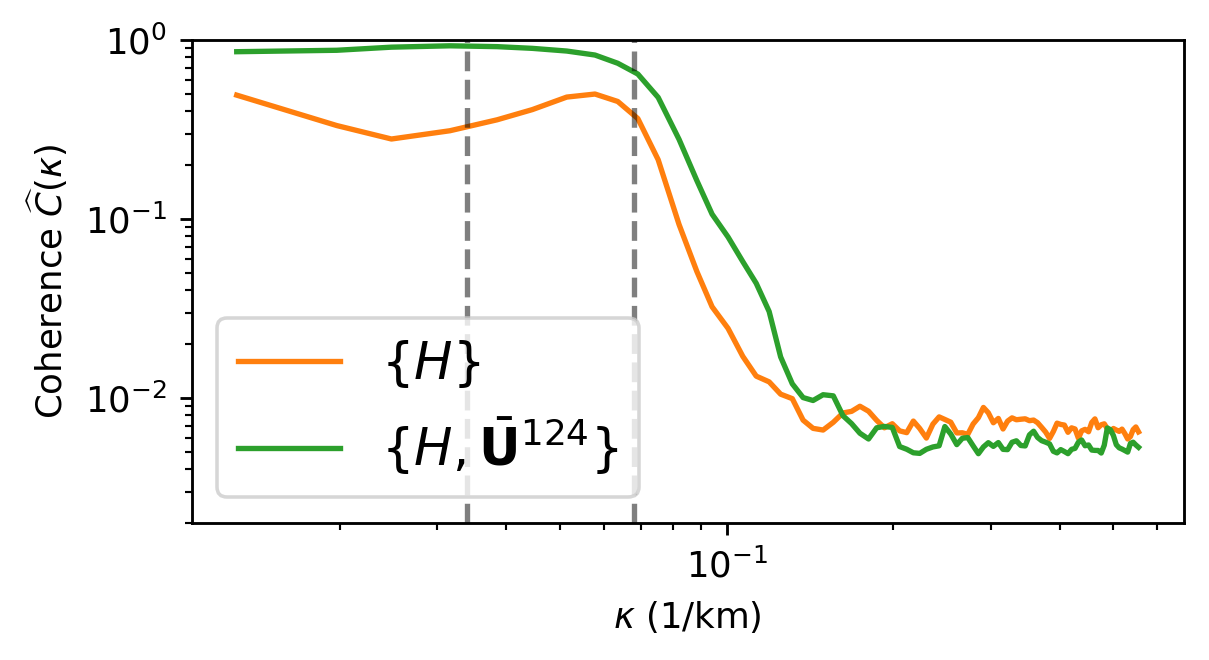}
\caption{\ch{The mid-jet coherence $\coh(\kappa)$ from Configuration $\{H\}$ [from Fig.~\ref{fig:coh_HvsHUT}, panel (a)] compared against the mid-jet $\coh(\kappa)$ from Configuration $\{H,\bar{\bs{U}}^{124}\}$ [from Fig.~\ref{fig:coh_filteredconfigs}, panel (c)].
}}
\label{fig:Coh_H_vs_HfiltU}
\end{figure}

\ch{We re-run each configuration involving $T$ or $\bs{U}$ after replacing the corresponding input field by its smoothed counterpart. Each configuration is retrained 10 times, and the variation across retrained U-Nets is small (Supporting Information Tables S3--S4). For brevity, Configuration $\{T\}$ is omitted because it already has negligible skill, and in configurations involving both $T$ and $\bs{U}$, only one of the two fields is smoothed at a time.
\\
We inspect the scale-dependent performance via the coherence $\coh(\kappa)$ [defined in \eqref{coherence}] in the mid-jet region [Fig.~\ref{fig:coh_filteredconfigs}, panels (c--i)]. To summarize the scales at which smoothing substantially degrades performance, we define a damping-scale wavenumber $\kappa_{0.5}$ as the smallest wavenumber at which the ratio
\beq \label{coherenceratio}
\frac{\text{$\coh(\kappa)$ from a smoothed-input configuration}}{\text{$\coh(\kappa)$ from the corresponding unsmoothed-input configuration}}
\eeq
first falls to or below $0.5$. The values of $\kappa_{0.5}$ are marked in Fig.~\ref{fig:coh_filteredconfigs}. In some smoothed-input configurations, the damping is too weak for the ratio \eqref{coherenceratio} to drop to $0.5$ anywhere; in these cases, $\kappa_{0.5}$ is not marked [panels (d,f,h) in Fig.~\ref{fig:coh_filteredconfigs}].
\\
The coherence $\coh(\kappa)$ from configurations involving SSV are shown in panels (c,e,g,i) of Fig.~\ref{fig:coh_filteredconfigs}. When $\bs{U}$ is smoothed, the $\coh(\kappa)$ consistently decreases at high wavenumbers $\kappa$. At the corresponding high $\kappa$ in the azimuthally averaged power spectra of the outputs (not shown), the spectra from smoothed-$\bs{U}$ configurations are also consistently lower than those from the corresponding unsmoothed-$\bs{U}$ configurations. Together, these behaviors indicate that smoothing $\bs{U}$ aggravates error type (iii), namely the blurring of small-scale features, discussed in \S\ref{sec:capturedmissed}.
Unsurprisingly, smoothing to an FS of 124 km aggravates this effect more than smoothing to an FS of 25 km; within each of panels (c,e,g,i), this is reflected by the stronger drop  in $\coh(\kappa)$ and the earlier (smaller) $\kappa_{0.5}$ in Configurations $\{\bar{U}^{124},\cdots\}$ than in Configurations $\{\bar{U}^{25},\cdots\}$, where ``$\cdots$'' denotes any additional input fields. 
\\
Comparing across panels (c,e,g,i)  of Fig.~\ref{fig:coh_filteredconfigs}, the damping-scale wavenumbers $\kappa_{0.5}$ are similar between the two configurations that include $H$ [panels (c,e)] and between the two configurations that do not include $H$ [panels (g,i)]. The with-$H$ configurations have later (larger) $\kappa_{0.5}$ than the without-$H$ configurations, suggesting that they are less sensitive to smoothing of $\bs{U}$. From our ``wave signature and scattering medium'' perspective (\S\ref{sec:wave_sig_scatter_med}), this is consistent with $H$ already supplying abundant wave-signature information, so that performance is less sensitive to the fine-scale information lost when $\bs{U}$ is smoothed.
\\
Smoothing $\bs{U}$ does not lead to a substantial ($>10\%$) decrease in coherence at spatial scales comparable to or larger than the incoming mode-1 IT wavelength, in any of the configurations [panels (c,e,g,i) of Fig.~\ref{fig:coh_filteredconfigs}], where the reference output fields are most energetic [panel (a) of Fig.~\ref{fig:Unet azim spectra}]. The behavior of $(\Upsilon,R^2)$, reported in Supporting Information Tables S3--S4, is consistent with this. In every configuration where $\bs{U}$ is smoothed, $\Upsilon$ and $R^2$ decrease somewhat, but in most cases the decrease remains small relative to the spread across retrained models. As a heuristic quantification, we compare the ensemble mean and standard deviation $\sigma$ of mid-jet $(\Upsilon,R^2)$ over the 10 retrained models for each configuration. The only smoothed-$\bs{U}$ configurations whose ensemble-mean mid-jet $(\Upsilon,R^2)$ decrease by more than $3\sigma$ relative to their unsmoothed counterparts are the without-$H$ configurations in which $\bs{U}$ is smoothed to an FS of 124 km:
\begin{itemize}
    \item Configuration $\{\bar{\bs{U}}^{124},T\}$, where $(\Upsilon,R^2)$ decreases from $(0.92,0.85)$ to $(0.89,0.79)$, and
    \item Configuration $\{\bar{\bs{U}}^{124}\}$, where $(\Upsilon,R^2)$ decreases from $(0.90,0.81)$ to $(0.87,0.76)$.
\end{itemize}
In other words, in all the with-$H$ configurations, and in all the configurations where $\bs{U}$ is smoothed only to an FS of 25 km, smoothing $\bs{U}$ does not appear to change $(\Upsilon,R^2)$ beyond retraining variability. Even in Configurations $\{\bar{\bs{U}}^{124},T\}$ and $\{\bar{\bs{U}}^{124}\}$, the decreases in $(\Upsilon,R^2)$ remain moderate: these configurations still substantially outperform Configurations $\{H\}$ and $\{T\}$, whose mid-jet $(\Upsilon,R^2)$ are $(0.68,0.46)$ and $(0.03,-0.03)$, respectively. Thus, smoothing $\bs{U}$ mainly degrades performance at smaller spatial scales, while having only limited impact on $(\Upsilon,R^2)$ and on performance near the dominant IT scales. This suggests that smoothing $\bs{U}$ does not strongly aggravate errors of types (i) and (ii) in \S\ref{sec:capturedmissed}, which are tied to dominant-scale structure and overall amplitude.
\\
Comparing the coherence from Configuration $\{H\}$ with that from Configuration $\{H,\bar{\bs{U}}^{124}\}$ (Fig.~\ref{fig:Coh_H_vs_HfiltU}), the improvement from adding $\bar{\bs{U}}^{124}$ spans a broad range of spatial scales and extends to output wavenumbers beyond those at which $\bar{\bs{U}}^{124}$ retains substantial spectral energy [see Fig.~\ref{fig:coh_filteredconfigs}, panel (a)]. For example, at $\kappa = 2\pi/(63)\,\mathrm{km}^{-1}$, the spectrum of $\bar{U}^{124}$ is strongly damped, being more than $10^5$ times smaller than that of the unfiltered $\bs{U}$, while the coherence from Configuration $\{H,\bar{\bs{U}}^{124}\}$ is still 3.2 times as large as that from Configuration $\{H\}$. 
This suggests that coarse-resolution SSV can still improve the reconstruction of IT patterns at output scales finer than the active scales in the input.
To leading order, wave signatures associated with a given horizontal wavelength project onto the same horizontal scales in $\bs{U}$ and $H$. The persistence of the improvement beyond the active scales of $\bar{\bs{U}}^{124}$ suggests that information other than direct fine-scale wave signature is useful, and $\bs{U}$ is used in a cross-scale manner by the U-Net. 
Speculatively, the U-Net may learn, to some degree, how BM-related $\bs{U}$ modulates (through refraction and scattering of) the finer-scale IT patterns in $H$, utilizing the larger-scale information on scattering media contained in the $\bs{U}$ to improve the finer-scale patterns.}

\ch{Taken together, these results suggest that smoothing $\bs{U}$ weakens performance at small spatial scales. Nevertheless, even strongly smoothed $\bs{U}$ still improves the reconstruction, including at small spatial scales, relative to omitting $\bs{U}$ altogether. This supports the view that $\bs{U}$ contributes not only direct wave-signature information, but also larger-scale contextual information that remains useful after smoothing.
\\
The coherences $\coh(\kappa)$ from configurations involving SST are shown in panels (d,f,h) of Fig.~\ref{fig:coh_filteredconfigs}. When $T$ is smoothed,  $\coh(\kappa)$ decreases perceptibly only in Configuration $\{H,\bar{T}^{124}\}$. In all other cases, the coherence ratio \eqref{coherenceratio} never falls to or below $50\%$. The statistics of $(\Upsilon,R^2)$ in Supporting Information Tables S3--S4 give a similar story. The only smoothed-$T$ configurations whose ensemble-mean mid-jet $(\Upsilon,R^2)$ are lower than those of their unsmoothed counterparts by more than $3\sigma$ are Configuration $\{H,\bar{T}^{124}\}$, where $(\Upsilon,R^2)$ decreases from $(0.79,0.62)$ to $(0.76,0.58)$, and Configuration $\{\bs{U},\bar{T}^{124}\}$, where $(\Upsilon,R^2)$ decreases from $(0.92,0.85)$ to $(0.91,0.82)$. These decreases remain moderate, and both configurations still outperform their corresponding counterparts without $T$.
These modest performance decreases are consistent with our interpretation that, in this dataset, $T$ mainly provides BM-related contextual information, which is concentrated at relatively large spatial scales and is therefore less sensitive to smoothing (\S\ref{sec:wave_sig_scatter_med}).
\subsection{Temporally misaligned inputs}
In realistic observing systems, $H$ may be measured by a different platform from $T$ or $\bs{U}$, so the available input fields need not be simultaneous. To probe the impact of such temporal misalignment, we run Configurations $\{H(t),\bs{U}(t-48\,\mathrm{h})\}$ and $\{H(t),T(t-48\,\mathrm{h})\}$, 
where $H$ leads $\bs{U}$ or $T$ by 48 hours (4 IT periods). The target outputs remain the IT imprints in $H$ at time $t$. We retrain these configurations 10 times and again find little variation across retrained U-Nets. We also ran the opposite offsets, $\{H(t),\bs{U}(t+48\,\mathrm{h})\}$ and $\{H(t),T(t+48\,\mathrm{h})\}$, which show similar performance (Supporting Information Tables S5--S6), and are skipped here for brevity.
Both temporally misaligned configurations perform worse than their simultaneous counterparts. The mean mid-jet $\Upsilon$ decreases from $0.94$ in Configuration $\{H,\bs{U}\}$ to $0.82$ in Configuration $\{H(t),\bs{U}(t-48\,\mathrm{h})\}$, and from $0.79$ in Configuration $\{H,T\}$ to $0.73$ in Configuration $\{H(t),T(t-48\,\mathrm{h})\}$. Nevertheless, both temporally misaligned configurations still outperform Configuration $\{H\}$, whose mean mid-jet $\Upsilon$ is $0.68$. This suggests that $\bs{U}$ or $T$ remain beneficial even when shifted relative to $H$ by a few IT periods.
\\
As the time shift equals a fixed multiple of IT periods in our setup, this experiment does not impose an arbitrary phase mismatch for IT signals that are coherent over multiples of $4$ IT periods.
Thus, the U-Nets may exploit residual phase relationships associated with the more coherent part of the IT field. 
Another plausible mechanism is that a substantial part of the BM field evolves on timescales longer than 48 h, so some BM-related contextual information remains correlated across the offset and can still be utilized by the U-Net.
Under both mechanisms, conceptually, it is crucial that the time shift be kept fixed during training and testing. 
Operationally, when observations of $\bs{U}$ or $T$ are not simultaneous with $H$, we suggest training the U-Net with inputs separated by the same temporal gap as in the target observations.
Both mechanisms also rely on the time gap not being too large. As the gap increases, less BM-related contextual information persists between snapshots, and fewer IT components remain coherent across the offset, leaving less usable wave-signature information. The 48 h offset considered here is a moderate mismatch test rather than an extreme one. Presumably, the performance would decrease further for larger time gaps. We leave a systematic study of this dependence to future work, to be motivated by specific observing applications.
\\
Taken together, these experiments suggest that the utility of $\bs{U}$ is not restricted to the idealized case of perfectly simultaneous, high-resolution inputs: in our simulation-based setting, coarsening and temporal misalignment both degrade performance, but $\bs{U}$ remains beneficial relative to using $H$ alone. 
}

\section{\ch{Impact of Non-locality}} \label{sec:nonlocal}
\begin{figure}
    \centering
    \includegraphics[width=0.9\linewidth]{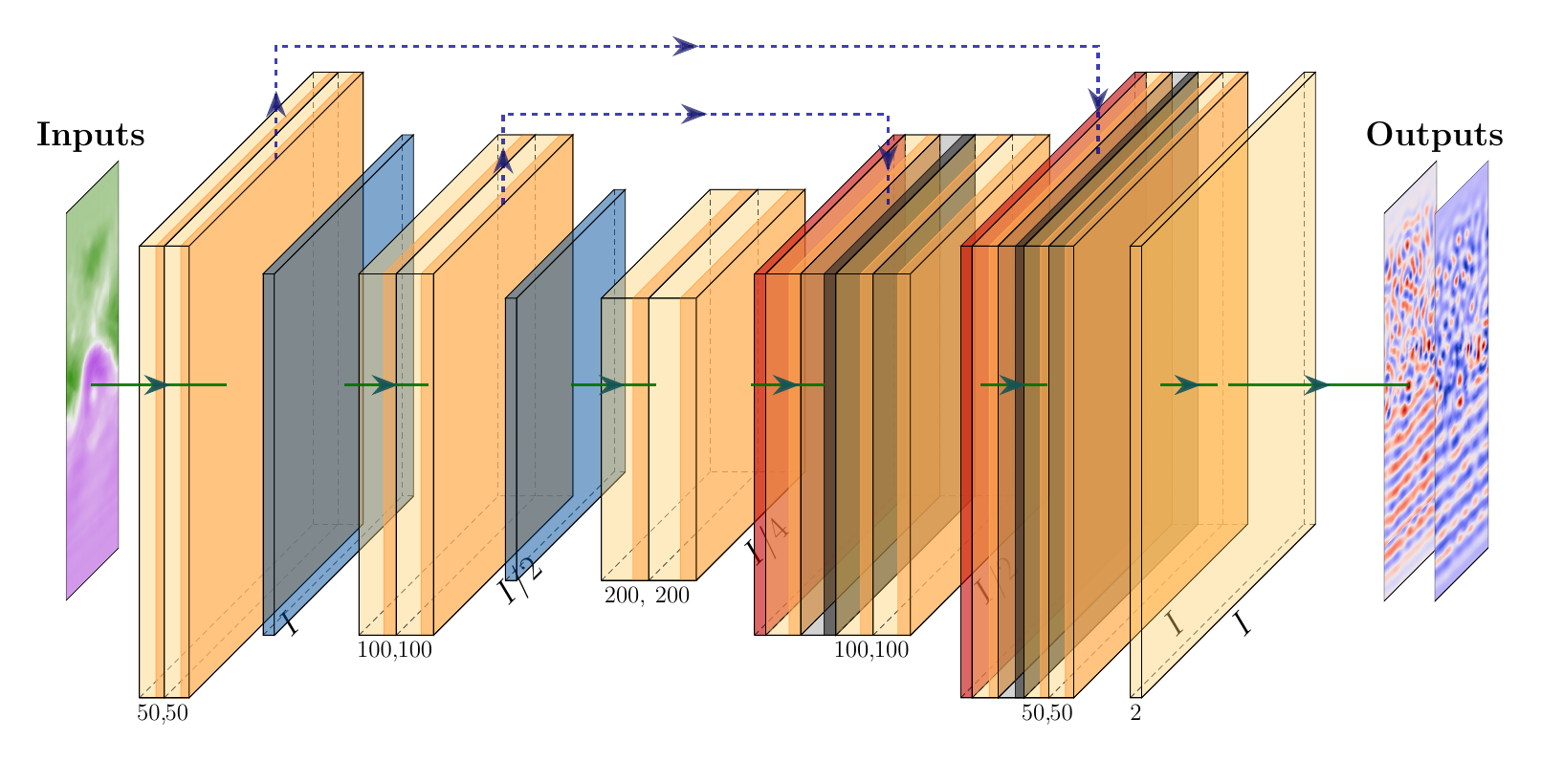}
    \caption{Architecture of the shallow U-Net with two encoding/decoding layers, used to test the role of nonlocal information. The labels are the same as in Fig.~\ref{fig:Unet architecture}. Further implementation details are provided in \citeA{han_wang_2026_19829961}.} 
    \label{fig:sketch_shallowUNet}
\end{figure}

The background conditions and BMs that comprise the scattering medium can be active at spatial scales much larger than the ITs' wavelengths.
\ch{Thus, an optimally performing IT extraction algorithm can reasonably be expected to benefit from access to spatial context extending beyond the IT scales. In our experiments with smoothed $\bs{U}$ as inputs (\S \ref{sec:degraded}), we find that the large-scale information in $\bs{U}$ is clearly harnessed. 
To harness information at large scales, neural networks of choice must be able to access non-local information. 
In convolutional neural networks, such access is enabled by a sufficiently  large receptive field, whose size depends on architectural choices such as depth, downsampling, kernel size, and dilation. In our main U-Net, nonlocal information is primarily enabled by the four encoding--decoding steps described in \S\ref{sec:dl_algorithm}.
\\
For our main U-Net defined in \S \ref{sec:dl_algorithm} with four encoding–decoding steps, the theoretical receptive field, i.e., the spatial span of the region of the input image that \ch{can influence} a single grid point in the output, is about $200\times 200$ grid points (see e.g., \citeA{araujo2019computing} for the computation of theoretical receptive fields).
With a grid spacing of $4$ km, this corresponds to a spatial span of about $800$ km in each horizontal direction, which is comparable to the meander width of the turbulent jet. 
\\
In practice, trained convolutional neural networks often use only a fraction of their theoretical receptive field \cite{luo2016understanding}. We therefore also estimate a gradient-based effective receptive field (ERF) under Configuration $\{H\}$. For each probed output location, we define a scalar probe equal to the U-Net output at that location and backpropagate to obtain the gradient of that probe with respect to each input pixel. We take the absolute value of the gradient as a proxy for local sensitivity. 
The probed output locations lie on a coarse lattice with spacing of $5$ pixels in both $x$ and $y$. For each such location, the sensitivity map (gradient amplitudes) is computed separately for the two output channels and then averaged over the two channels and over $30$ test snapshots.
Our choice to take the absolute values before averaging is intended to avoid sign cancellations, as our output channels represent different phases of the oscillatory IT fields. From the resulting  averaged sensitivity map for each probed output location, we define $r_{95}$ for each probed output as the smallest radius such that the cumulative gradient magnitude within a circle of radius $r_{95}$ centered on the probe location reaches $95\%$ of the total. This is analogous to an encircled-energy radius in optics \cite{smith2000modern}. A notebook showing these calculations is provided in \citeA{han_wang_2026_19829961}. In the mid-jet region, the mean $r_{95}$, averaged over all probed output locations on the lattice, is $78$ grid points, or $312$ km. Although this ERF measure is heuristic, as it is based on gradients and therefore probes local, infinitesimal sensitivity, neglecting responses to, say,  finite-amplitude perturbations, it nevertheless suggests that the trained network uses nonlocal information in practice.
\\
To test more directly whether such access to nonlocal information matters, we construct a shallow U-Net with only two encoding--decoding steps (Fig.~\ref{fig:sketch_shallowUNet}) and consider two variants under Configuration $\{H\}$. The first variant (``no dilation'') uses the same setting of convolutional kernels as the main U-Net at the corresponding layers. The second variant (``with dilation'') enlarges the theoretical receptive field by replacing the two $3\times3$ convolutions in the bottleneck with dilated $3\times3$ convolutions with dilation rates of $2$ and $3$ (and with padding chosen to preserve feature-map sizes). The two shallow variants have the same depth and nearly identical parameter counts, differing mainly in the spatial extent over which bottleneck features can aggregate information.  Because of this, we regard their comparison as a clean test on the role of nonlocality.
\\
The shallow U-Net without dilation has a theoretical receptive field of $44\times 44$ grid points, corresponding to a span of $176$ km in each horizontal direction. The shallow U-Net with dilation has a theoretical receptive field of $68\times 68$ grid points, corresponding to a span of $272$ km in each horizontal direction. To avoid introducing a large change in model capacity, we increase the number of kernels in the shallow U-Nets so that both shallow variants have a similar number of trainable parameters to that of the main U-Net; each shallow variant has $4.5\%$ more parameters than the main U-Net. All other aspects of training are kept the same. Each shallow U-Net variant is retrained five times, and the spread across retrained U-Nets is small (Supporting Information Tables S7--S8). In principle, an even more local architecture (e.g., with fewer downsampling stages and thus larger feature maps at the bottleneck) would provide a more illustrative ``local-only'' baseline; this is currently computationally prohibitive, due to the larger feature maps that exceed the available GPU memory, if we are also to keep the number of trainable parameters comparable.  
\\
For brevity, we only inspect mid-jet $\Upsilon$ and $R^2$, which are summarized in Table~\ref{tab:shallowUNet}. The shallow U-Net with dilation performs substantially better than the shallow U-Net without dilation, while remaining only slightly weaker than the main U-Net (also reflected in the statistics of retrained U-Nets in Tables S7--S8 and S1--S2). As the two shallow variants have the same depth and nearly identical parameter counts, this comparison provides evidence that enlarging the accessible spatial context is beneficial. In particular, a spatial span of $176$ km in each horizontal direction appears insufficient, whereas enlarging the field of view to $272$ km substantially recovers the lost skill. 
\\
This dependence on receptive-field size is problem-specific. For example, in a different context, learning subgrid closures for two-dimensional turbulence, \citeA{srinivasan2024turbulence} finds that performance is optimal for a much smaller theoretical receptive field of $9\times 9$ grid points.  The relevant level of non-locality is evidently much shorter there than in our IT-BM disentanglement problem.  
Some non-locality is still essential in the turbulence closure problem, as shown by \citeA{brolly2025stochastic}; this is consistent with \citeA{srinivasan2024turbulence}, which argues that small-scale fluxes are physically related to neighboring grid points via (for example) terms involving spatial gradients, and weak non-locality still matters. In our problem, as the scattering medium is active at spatial scales much larger than the IT wavelengths, it makes sense physically that large, mesoscale-reaching receptive fields are helpful.
\\
In applications to satellite observations, these results suggest that wave--mean disentanglement should, in principle, make use of broad spatial context, for example from wide-swath observations or long contiguous track segments. This contrasts with approaches that infer BMs from altimetry through geostrophic or higher-order balances, whose underlying dynamical relations are local in the sense that the inferred quantities depend only on SSH and its spatial derivatives \cite{tranchant2025swot,bertrand2025robust}. Such approaches do not (yet) use properties of wave signals.}

\begin{table}
\caption{\ch{Mid-jet $\Upsilon$ and $\ch{R^2}$ under Configuration $\{H\}$ for the main U-Net, the shallow U-Net without dilation, and the shallow U-Net with dilation. }}
\centering
\begin{tabular}{l|c|c|c}
\hline
Metric & Main & Shallow, no dilation & Shallow, with dilation \\
\hline
$\Upsilon \times 100$ & 68 & 64 & 68 \\
$R^2 \times 100$ & 46 & 39 & 44 \\
\hline
\end{tabular}
\label{tab:shallowUNet}\end{table}

\section{Summary and discussions} \label{sec:Unet:discussion}
In this study, we have systematically evaluated the ability of a deep learning model---specifically a \chm{deterministic, point-estimating} U-Net architecture---to extract IT signatures from different combinations of sea surface fields in an idealized Boussinesq simulation. We show that the U-Net's performance depends critically on the physical information present in the inputs. A clear hierarchy of performance among the different input configurations is established \chm{in the present deterministic benchmark}, with the surface velocity field, $\bs{U}$, providing by far the most significant contribution to an accurate separation. 

The relative success of each input field can be understood through the lens of the information they provide regarding the ``wave signature" and the ``scattering medium."

\begin{itemize}
\item Surface velocity ($\bs{U}$) \ch{emerges as} the most \ch{informative} input. It contains a strong signature of the IT's wave kinematics, as well as signatures from BM velocities that scatter ITs. 

\item Sea surface height ($H$) performs reasonably well. It contains the target wave signature by definition, but this signature is entangled with the much larger signal of the BM.

\item Surface temperature ($T$) contains only a weak IT signature but provides valuable, unambiguous information on the scattering medium, as it is  strongly related to the BM that scatters the waves. Adding $T$ to other inputs consistently improves performance by helping the network disambiguate the BM imprints from the total signal. 
\end{itemize}


The full combination of SSH ($H$), velocity ($\bs{U}$), and temperature ($T$) yields the best performance, achieving a coefficient of determination ($\ch{R^2}$) of 0.95 and a correlation ($\Upsilon$) of 0.97 against the reference fields. Residual errors (the remaining 5\% of unexplained variance) concentrate at small spatial scales. 
These residual errors 
\chm{likely reflect a combination of three factors: contamination in the reference fields (from, e.g., Doppler-shifted wave signals), architectural limitations of the present U-Net, and the possibility that some of the small-scale imprints  are only weakly constrained by the input information. }

A central finding of this work is that surface velocity is the single most 
\chm{useful observable in the present deterministic framework}
for IT-BM separation. Intriguingly, \citeA{He2024}, who focus on inferring interior vertical velocities, reach a similar conclusion regarding the importance of surface velocities. This apparent agreement on the critical role of surface velocities strongly advocates for the development and deployment of future satellite missions capable of measuring sea surface vector velocities, such as  \st{ODYSEA, }SEASTAR and HARMONY. 

Our results also highlight the immense value of coordinating multi-platform observational campaigns. For instance, aligning measurements from ship-based measurements, high-frequency radar, or surface drifters with SWOT and SST satellite tracks could achieve the synergistic benefits demonstrated in our $\{H, \bs{U}, T\}$ configuration. Similar discoveries have been made for other purposes too. For example, the \ch{synergistic} benefit from combining SST and SSH observations is explicitly demonstrated when machine learning \cite{Martin2023} or data assimilation methods \cite{le2025vardyn} are applied to produce gap-free SSH (and SST) data.

Our investigations provide some broader lessons on neural networks:
\begin{itemize}
    \item Architecture choices. Despite its relative simplicity, the U-Net performs on par with the more complex cGAN used in W22. The crucial element for the U-Net's success is a learning rate that varies during training. 
    The computational efficiency of U-Nets allows us to run many different experiments conveniently. Moreover, the architectural simplicity and the stable training behavior of U-Nets allow us to understand more clearly what factors are conducive to its success. For other applications, a U-Net with a varying learning rate can be a \ch{useful} baseline algorithm.  
    \item  Physical interpretability. The ``wave signature vs. scattering medium" framework provides a clear lens for understanding why some inputs perform better than others. This perspective is transferable to other studies.
    \item Non-locality. \ch{Access to mesoscale-reaching spatial context is important for the U-Net's success in our application}, consistent with the physical understanding that IT–BM interactions take place at large spatial scales. For other applications, the optimal receptive field depends on the problem at hand. 
\end{itemize}


In future works, the small-scale error can be mitigated \ch{through} (a) more careful data preparation or (b) changes in the deep learning architecture. For (a), more accurate methods to separate the ITs and BMs can prevent the physical imprints of ITs from intermingling with spurious signals in the reference fields, which currently confuse our U-Net. Examples include
time-filtering conducted in flow-following frames \cite{ShakespeareGibsonHogg2021, Kafiabad2022, baker2025lagrangian}, and filtering methods that do not rely on the separation in time scales \cite{Early2021, wang2023dynamical}.
For (b), \ch{as summarized in the Introduction, different approaches are attempted in several related works \cite{lyu2024multi,Wang2025MultiScale,liu2025wide}. A} dedicated comparison of small-scale behaviors between different architectures is required. \ch{In addition to the existing approaches, }we speculate that in particular, attention mechanisms can help address the localized, velocity-associated errors observed in Fig.~\ref{fig:height jet}. 

\chm{Even with such improvements, however, the present U-Net returns only a single best-guess reconstruction for each input snapshot, and does not quantify or disentangle sources of uncertainty. This makes the U-Net's errors harder to interpret. 
This could be addressed in two complementary directions \cite{kendall2017uncertainties}. First, probabilistic formulations that predict a distribution of plausible outputs, which have been explored in \citeA{Wang2025MultiScale} and 
in other ocean dynamics related applications  \cite{ foster2021probabilistic,clare2022explainable,brolly2023inferring}, would probe uncertainty when the inputs do not tightly constrain the output and avoid collapsing the possible outcomes into a single smoothed prediction. Second, uncertainty associated with model and training choices could be probed through approaches such as Monte-Carlo dropout \cite{gal2016dropout}. 
Distinguishing these sources of uncertainty would help clarify whether residual errors arise primarily from incomplete constraint by the available inputs or from limitations of the estimator.
}

Careful readers may notice that under Configuration $\{H\}$, the performance metrics ($\Upsilon, \ch{R^2}$) from our U-Net are slightly stronger than those of the cGAN in W22; this difference arises primarily from small changes in data division/selection (\S \ref{subsec:inputoutput} and Supporting Information Text S1). Qualitative differences between the U-Net and the cGAN do appear at small spatial scales (below the mode-1 tidal wavelengths): our U-Net tends to \ch{smooth} out the fine structures, whereas the cGAN in W22 tends to create random small-scale patterns that are \ch{often locally} incorrect, in an effort to mimic small-scale statistical behaviors.
\chm{This contrast suggests that part of the residual small-scale mismatch is dependent on the algorithm. This comparison further motivates the probabilistic formulations in future works.}

From Fig.~\ref{fig:Unet success and failure} and other snapshots (not shown), the U-Net---under any Configuration we tested---appears to correctly recognize that ITs are more coherent in the up-jet region, as indicated by its generation of patterns resembling plane waves there. The same behavior is evident for the cGAN in W22. This suggests that neural networks may be used not only for extracting ITs but also for identifying regions of strong incoherence. \chm{If such regions also tend to have less tightly constrained outputs, then probabilistic frameworks that quantify predictive uncertainty may help identify them.}

We hypothesize that an additional, particular advantage of Configuration $\{\bU\}$ is that, in our data, ITs are the dominant source of divergent surface motion, while the BM is largely rotational. Internal waves with intrinsic frequencies above the Coriolis frequency project strongly onto divergence (e.g., \citeA{buhler2014wave}), whereas balanced, quasi-geostrophic flows (such as the turbulent jets in our simulations) that constitute the scattering background project weakly on the divergent component (e.g., \citeA{gill2016atmosphere}). The snapshot shown in Fig.~ \ref{fig: vel and vort inputs} demonstrates this visually:  the divergence $D$ is  dominated by spatial scales around tidal wavelengths, with plane-wave-like patterns up-jet and scattered-wave-like patterns mid-/down-jet, consistent with the dominance of ITs, whereas the vorticity $\zeta$ visibly contains both imprints from ITs and large-scale structures contributed by BMs. 
Conceptually, this could provide the U-Net with a relatively clean pathway to distinguish (and hence use) the wave signature (in the divergence) from the imprints of scattering medium (largely in the vorticity). In some oceanic scenarios, unlike in our dataset, BMs can have strong imprints on divergent currents due to the presence of strong submesoscale currents typical of the winter mixed layer (e.g., \citeA{barkan2019role}). There, the disentanglement problem may be more complicated. 

Out of theoretical curiosity, we also experiment with configurations where $D$ and/or $\zeta$ are inputs. Supporting Information Text S4 reports details. Configuration $\{D\}$ outperforms any other configuration that has just one input channel, suggesting that the strong wave signatures in $D$ is useful. Configuration $\{\zeta\}$ exhibits problematic behavior (i.e., counterintuitive responses to some training parameters we do not observe in other configurations); moreover, Configuration $\{\zeta,D\}$ performs slightly worse than Configuration $\{\bU\}$, even though conceptually, the separation between waves and BMs in $(\zeta,D)$ should be more straightforward than $\bU$. These counter-intuitive behaviors suggest that our U-Net is not optimal for inferring information from $\zeta$ and/or $D$. This may be because $\zeta$ and $D$ are active at spatial scales smaller than $\bU$, and our U-Net may not focus sufficiently on extracting information from small scales. 

In addition to new observational campaigns, another promising approach to obtain surface current velocities is through data-driven algorithms taking existing observed fields as inputs. For example, \citeA{zhou2025machine} infer surface velocities from SWOT's SSH observations through deep learning; \citeA{fablet2024inversion} infer the evolution of surface velocities from SSH and SST via a combination of deep learning and data assimilation, harnessing the dynamical links between the surface fields; and \citeA{lenain2026unprecedented} infer surface velocities from  \ch{SST }and surface heat flux observations, harnessing the kinematic relationship between velocities and SST \ch{evolution governed by} the advection-diffusion equation. Such approaches can be aided by gap-free data products constructed from along-track satellite observations \cite{Martin2023,le2025vardyn}, which provide more continuous inputs for training and application.


We choose to extract the IT imprints on SSH, rather than the BM imprints. In our datasets, the latter task would produce artificially high performance metrics because the input and output would be highly similar: the imprints from the BMs dominate the total SSH signal, taking up more than $90\%$ of the total variance of SSH. In datasets where the BMs do not overwhelmingly dominate the signals (e.g., \citeA{Gao2024}), it can still be meaningful to evaluate performance metrics for BM extraction. 

\ch{
More generally, the image-to-image formulation used here is not restricted to IT imprints on SSH. 
If suitable reference labels are available, the output channels could instead be defined as IT-related velocity components, or diagnostics related to IT-induced mixing or dissipation. 
Such extensions could be scientifically valuable, as they would connect remotely observed surface fields more directly to the life cycles of IT energy. 
The skill of such reconstructions would need to be assessed separately, especially as velocity- and mixing-related targets may be more sensitive to small-scale and/or subsurface structures than the SSH imprint considered here. 
Such alternative-output formulations represent a promising future direction.
}

Our work and its counterparts primarily concern new-generation satellite observations, which provide two-dimensional spatial coverage but suffer from poor temporal sampling. Mooring data represent the opposite case: they lack spatial coverage but provide higher temporal resolution, typically with sampling intervals on the order of an hour---well below tidal periods.
Despite the superior temporal sampling, incoherence remains a major challenge for harmonic fitting of mooring data too, as phase shifts of ITs can still vary \ch{over the fitting windows used to estimate harmonic components}. Recent works show that deep learning---particularly encoder–decoder architectures similar in spirit to our U-Net---has promise for addressing this challenge in time series data as well \cite{li2023changing}.

\ch{At present, the wave/IT-BM disentanglement community still lacks a shared set of  evaluation diagnostics.
Our results suggest that future benchmark evaluations for snapshot-based BM--wave disentanglement should not rely solely on domain-averaged scalar skill metrics such as mean correlation or mean $R^2$. These remain useful summary diagnostics, but they can mask scale-dependent differences between methods. In our own experiments, for example in \S 5.1, spatial smoothing of input fields can substantially degrade performance at high wavenumbers while barely affecting the domain-averaged $(\Upsilon, R^2)$. Diagnostics in spectral space, such as power spectra and spectral coherence, are therefore important complements when assessing how well different methods recover physically relevant structures across scales. If particular scale ranges are of interest (for example, small scales around or below mode-2 IT wavelengths in our work), additional summary measures, such as a damping-scale metric like our $\kappa_{0.5}$, may also be useful. A community benchmark would ideally include both compact scalar scores and scale-aware diagnostics.}

While we provide a clear proof-of-concept using an idealized model, future work must progress toward more complex and realistic scenarios. The methodology should be tested on data from comprehensive general circulation models that include realistic bathymetry, a full spectrum of internal waves, and geographically varying stratification. The simpler, computationally cheaper deep learning algorithm identified in this work contributes to efforts in this direction. 
A critical future step will be to improve the reference datasets used for training, potentially by employing more sophisticated separation techniques. \ch{In this direction, in a closely related setting (wave-BM separation), \citeA{lyu2024multi} and \citeA{Wang2025MultiScale} both rely on a publicly available, regional data set where the labels are derived by Lagrangian filtering \cite{jones2023using}. Such labels can at least alleviate spurious signals associated with purely Eulerian frequency-based separation, including Doppler-shift-related contamination; at present, publicly available benchmark datasets with IT or wave labels derived from Lagrangian filtering appear limited. } 
Ultimately, the goal is to apply these trained models to actual satellite observations, a step that will require careful handling of instrument noise and data gaps. These explorations, guided by the findings herein, will be essential to fully leverage the wealth of new-era satellite data and make meaningful progress in understanding the ocean's intricate multiscale dynamics.

\section*{Open Research Section}
Production codes used for this paper are available on Zenodo \cite{han_wang_2026_19829961}. This includes scripts for the definitions and the training of our U-Net architecture, and for the analysis of outputs. Potential future updates of the codes can also be accessed on GitHub \url{https://github.com/hannnwang/Extract_internal_tides_with_UNet_and_surface_field_synergy}. The data used for the training and testing of our U-Net are published on Scholars Portal Dataverse \cite{BoussData}. 

\section*{Conflicts of Interest}
The authors declare there are no conflicts of interest for this manuscript.

\acknowledgments
This paper is a contribution to the projects W2, L2 and M2 of the Collaborative Research Centre TRR 181 ``Energy Transfers in Atmosphere and Ocean" funded by the Deutsche Forschungsgemeinschaft (DFG, German Research Foundation) - Projektnummer 274762653, which supports HW. 
J.U.\ and N.G.\ acknowledge the support of the Canadian Space Agency [14SUSWOTTO] and of the Natural Sciences and Engineering Research Council of Canada (NSERC) [RGPIN-2022-04560]. K.S. was supported by the Office of Naval Research (N00014-25-1-2183).
\ch{We acknowledge financial support from the Open Access Publication Fund of Universität Hamburg.}

We thank Julien Le Sommer, Callum Shakespeare, No\'e Lahaye and Belal Abdelhadi for helpful discussions.  
Simin Wang produced Supporting Information Fig.~S1 as part of her master thesis at University of Edinburgh. We thank Aur\'elien Ponte for helpful discussions, and for supplying simulation output data used in W22 and reused here. \ch{Comments from Martin Brolly and two anonymous reviewers have significantly improved the manuscript. }


\end{document}